\documentclass[aps,prd,twocolumn,groupedaddress,floatfix,showkeys,nofootinbib]{revtex4-1}

\usepackage{amsmath}   
\usepackage{amsfonts}  
\usepackage{amssymb}
\usepackage{graphicx}
\usepackage{dcolumn}
\usepackage{bm}
\usepackage{natbib}
\usepackage{url} 
\usepackage{hyperref}
\usepackage{subfigure}
\usepackage{cases}
\newcommand{\subsubsubsection}[1]{\paragraph{#1}\mbox{}\\}
\newenvironment{rcases}
  {\left.\begin{aligned}}
  {\end{aligned}\right\rbrace}

\begin{document}

\title{Constraining the halo size from possible density profiles of hydrogen gas of Milky Way Galaxy }

\author{Sayan Biswas}
\email[]{sayan@rri.res.in}
\affiliation{Raman Research Institute, C.V. Raman Avenue, Sadashivanagar, Bangalore,  560080, India}

\author{Nayantara Gupta}
\email[]{nayan@rri.res.in}
\affiliation{Raman Research Institute, C.V. Raman Avenue, Sadashivanagar, Bangalore, 560080, India}

\begin{abstract}
Galactic magnetic field (GMF) and secondary cosmic rays (CRs) (e.g. $^{10}$beryllium, boron, antiproton) are important components to understand the propagation of CRs in the Milky Way Galaxy.  Realistic modeling of GMF is based on the Faraday rotation measurements of various Galactic and extragalactic radio  sources and synchrotron emission from CR leptons in the radio frequency range, thereby providing information of halo height. On the other hand,  diffusion coefficient and halo height are also estimated from the $^{10}$Be/$^{9}$Be and B/C ratios. Moreover, density distribution of gaseous components of interstellar medium (ISM) also plays an important role as secondary CRs are produced due to interaction of primary CRs with the gaseous components of ISM. We consider mainly molecular, atomic, and ionized components of hydrogen gas for our study. Recent observations and hydrodynamical simulations provide new forms of density profiles of hydrogen gas in Milky Way Galaxy. In the \texttt{DRAGON} code, we have implemented our chosen density profiles, based on realistic observations in radio, X-ray and $\gamma$-ray wavebands, and hydrodynamical simulations of interstellar hydrogen gas to study the variation in the height of the halo  required to fit the observed CR spectra. Our results show the halo height ($z_{t}$) varies in the range of 2 to 6 kpc for the density profiles considered in our work.
\end{abstract}

\keywords{cosmic rays, Galactic magnetic field, interstellar medium, \texttt{DRAGON} code.}

\maketitle

\section{\label{sec:intro}Introduction}
\noindent The study of the origin of Galactic cosmic rays (CRs) is one of the most enigmatic areas of research at present. Many theoretical models, starting from very simple version to complex ones, have been used  in past several decades to explain observational results successfully. Ginzburg and Syrovatskii \cite{ginzberg64} was the first who had introduced various terms in the transport equation, used to study propagation of CRs, to include possible gain and losses in the flux of CRs. Previously, leaky box model and its variants were widely used to study the observed GeV flux ratio of secondary to primary \cite{shapiro70, cowsik73}. The secondary CRs of lower atomic number  ($Z$) are mainly produced due to the interaction of primary CRs with matter of interstellar medium (ISM) and background radiation. Later, more complex models of CR propagation are evolved which include the energy dependence of the diffusion coefficient and re-acceleration scenario \cite{gupta89,gaisser91, berezinskii90, letaw93}. In recent developments, along with energy dependence, the spatial dependence of diffusion coefficient and more realistic structures of the Galactic magnetic field (GMF) have also been taken into account to solve the transport equation of CRs \cite{bernardo10,bernardo13,evoli17,cholis12,tomassetti16,gaggero14}. Despite such advancements, several important questions on CR propagation remain open. 

An important source of uncertainty in case of CR propagtion is the large error on halo height or the vertical height (perpendicular distance measured from Galactic plane) of CR  diffusion region which is in the range of $1$ to $10$ kpc as estimated  from $^{10}$Be/$^{9}$Be (Be stands for beryllium) ratio \cite{strong07}. A stronger constraint on the halo height i.e. $5.4 \pm 1.4$~kpc was obtained from the global analysis of the available data within a numerical propagation framework \cite{trotta11} while weaker constriants ($8^{+8}_{-7}$~kpc) were also estimated by following the same kind of analysis procedure with semianalytic codes \cite{putze10}. However, the effect of possible density profiles of hydrogen gas of Milky Way Galaxy on halo height was left out in those analysis. The aim of our present work is to study such effects. For the present study, we mainly need a realistic model of  GMF, $^{10}$Be/$^{9}$Be and B/C (B and C, here, denote boron and carbon, respectively) and different density profiles of hydrogen gas of Milky Way Galaxy.

GMF directly affects the diffusion of CRs in the ISM by deflecting the CRs, thereby erasing most of the angular information of the source and spatial distribution of Galactic CRs. Hence, more realistc description of z-dependent (z is the vertical distance from the Galactic plane) GMF model is needed to understand CR propagation and the synchrotron emission of angular distribution \cite{bernardo13}. Recent studies indicate that such realistic structure of GMF can be modelled from the Faraday rotation measurements \cite{han09,pshirkov11,jansson12} of a large number of Galactic and extragalactic radio sources along with the synchrotron emission from Galactic CR electrons in the radio frequency range \cite{bernardo13}. Hence, such GMF models would provide us a relation between GMF (especially the random/turbulent component) and halo height. Moreover, $^{10}$Be/$^{9}$Be and B/C are also important for our present study as  $^{10}{\rm{Be}}/^{9}{\rm{Be }}\propto \sqrt{D(E_{k})}/z_{t}$  and  ${\rm{B/C}} \propto z_{t}/D(E_{k})$ with $z_{t}$, $E_{k}$ and $D({E_{k}})$ being the halo height, kinetic energy and energy dependent diffusion coefficient \cite{strong11}. Due to weak dependence  on diffusion coefficient,  $^{10}$Be/$^{9}$Be would be considered as a possible probe for halo height (consistent with GMF) \cite{strong07,berezinskii90}. B/C ratio, on the other hand, would provide us information of diffusion coefficient.

In the present work, we need the knowledge of interstellar matter of Milky Way  to choose  different density profiles of hydrogen gas. The interstellar matter is distributed inhomogeneously, atleast at small scales, through out the Milky Way Galaxy and its major concentration is observed near the Galactic plane and along the spiral arms of the Galaxy. It has been estimated that interstellar matter contributes about $\sim 10-15\%$ of the total mass of the Galactic disk \cite{ferriere01}. The total mass of the ISM is distributed in the discrete clouds and the regions in between such discrete clouds.  The discrete interstellar clouds occupy roughly $\sim 1-2 \%$ of the total interstellar volume \cite{ferriere01}. Such discrete interstellar clouds can be divided into three types namely dark, diffuse and translucent clouds. The dark clouds essentially contain very cold molecular gas having temperature, $T \sim 10-20$~K which can block off the starlight coming from the background stars \cite{ferriere01}. The diffuse clouds consist of cold atomic gas having $T \sim 100$~K. Diffuse clouds mostly act as a transparent medium for the background starlight. But in specific wavelengths, diffuse clouds may produce absorption lines. Both molecular and atomic gases are the basic ingredients of the translucent clouds. The visual extinction is intermediate, i.e. in between of dark and diffuse clouds, in case of such clouds. The interstellar mass in between the regions of the clouds corresponds to three different forms of matter, namely warm (primarily neutral) atomic, warm ionized, and hot ionized. Here `warm' and `hot' refer to $T \sim 10^{4}~K$ and $T \sim 10^{6}~K$ respectively. Apart from those aforesaid  components, also there is interstellar dust in ISM. In the present study, we solely concentrate on the gaseous components of hydrogen in ISM of Milky Way Galaxy.

In 1930, interstellar molecules such as CH, $\rm{CH^{+}}$ and CN were discovered from the observation of optical absorption lines in stellar spectra produced by such molecules \cite{ferriere01}. However, until 1970, we did not even know the existence of the most abundant molecule, i.e, molecular hydrogen ($\rm{H_{2}}$), of ISM. In 1970, the ultraviolet (UV) astronomy from above the Earth's atmosphere has provided the first glimpse of $\rm{H_{2}}$ in far UV spectrum of a hot star \cite{carruthers1970} and it also opened up a new window to observe the Universe in a different wavelength.  CO, after $\rm{H_{2}}$, is another most abundant molecule in the ISM and it was discovered in the following year. The first major survey of interstellar molecular gas was done by the UV spectrometer installed on the \textit{Copernicus satellite} \cite{spitzer1975}. Despite the success of optical and UV astronomy, they suffer a major drawback of interstellar extinction. Actually, the optical and UV absorption lines are not useful for astronomers to study the interior of the dense molecular clouds as the interstellar dust present in those regions obscure the bright source to form absorption lines. This problem was overcome due to remarkable advancement of radio astronomy as the radio wavelengths do not face any interstellar extinction. Later, X-ray and $\gamma$-ray astronomy expand our observational field of view. Recent high resolution X-ray and $\gamma$-ray space telescopes along with hydrodynamical simulations help us to achieve important knowledge and proper mapping of interstellar material.

 Here, we study the variation in the halo height of Milky Way Galaxy  required to fit the CR data such as $^{10}$Be/$^{9}$Be, B/C, proton, helium and antiproton using the Diffusion Reacceleration and Advection of Galactic cosmic rays: an Open New code \texttt{DRAGON}\footnote{\url{https://github.com/cosmicrays/DRAGON}} \cite{bernardo10} by considering possible  density profiles of hydrogen gas (i.e., molecular, atomic/neutral and ionized), based on radio, X-ray and $\gamma$-ray observations, and different hydrodynamical simulations including the gas flow dynamics and the cosmological parameter \cite{ferriere1998,ferriere07,feldmann13,miller15,troitsky17}. It is to be noted that we, here, ignore the primary components of boron and antiproton and focus only on the secondary components produced due to interaction of primary CRs with the interstellar matter.

We organize our paper in the following way.  Sec. II contains the major components of interstellar matter of our Galaxy. In Sec. III, we discuss about the model parameters required to fit the observed data set of $^{10}$Be/$^{9}$Be, B/C, proton, helium and antiproton using the \texttt{DRAGON} code.  The following section, Sec. IV, is dedicated to the possible density profiles of hydrogen gas obtained from realistic observations and hydrodynamical simulations. In Sec. V, CR spectra consistent with observed data are obtained for the various density profiles of interstellar gas. The corresponding propagation parameters for each case are also provided in tabular form. Finally, we draw conclusion in Sec. VI.

\section{\label{sec:ism}Interstellar matter}
The chemical composition of interstellar matter has a close similarity with the composition of CRs. This is inferred on the basis of the abundance measurements in the Sun, in several stars and in meteorites. The interstellar matter consists of 90.8\% in number [70.4\% of mass] of hydrogen, 9.1\% [28.1\%] of helium, and 0.12\% [1.5\%] of heavier elements, commonly known as metals in the astrophysical community \cite{ferriere01}. In the following subsection, we briefly discuss about the major components (needed for our present work) and their detection techniques.

\subsection{Molecular gas}

Molecular hydrogen (${\rm{H_{2}}}$) is the most abundant molecule in the ISM of Milky Way Galaxy. The second most abundant molecule is CO. To study ${\rm{H_{2}}}$, radio observations are preferred than UV and optical observations as UV and optical wavelengths suffer interstellar extinction. However, the most surprising fact is that ${\rm{H_{2}}}$ does not have permitted transition in the domain of radio frequency as ${\rm{H_{2}}}$ has small moment of inertia with no permanent electric dipole moment \cite{ferriere01}. Hence, ${\rm{H_{2}}}$ is studied indirectly using the radio observations of CO molecules. CO molecule has ($J = 1 \rightarrow 0$) a rotational transition at a radio wavelength of 2.6 mm \cite{ferriere01}. Actually such transition of CO acts as a tracer of  ${\rm{H_{2}}}$, where, ${\rm{H_{2}}}$-to-CO conversion factor, known as $X_{CO}$, is used to obtain the information of ${\rm{H_{2}}}$.

The first large scale survey of CO with 2.6 mm emission was carried out by Scoville and Solomon \cite{scoville1975} and by Burton \textit{et al.} \cite{burton1975}. The surveys showed that ${\rm{H_{2}}}$ mostly resides within a ring extending radially, from the Galactic center, between $3.5$~kpc and $7$~kpc. Furthermore, it may be noted that a strong molecular concentration was also observed within $0.4$~kpc \footnote{These surveys assumed that the Sun is at a distance from the Galactic center, $R_{\odot} = 10$~kpc. Later, the distance of the Sun, on the recommendation of IAU, was changed to $R_{\odot} = 8.5$~kpc. So, the lengths are scaled down by a factor $0.85$.}. From the vertical point of view, ${\rm{H_{2}}}$ concentrates mostly near the Galactic plane. The vertical distribution of ${\rm{H_{2}}}$ is assumed to be Gaussian for modelling \cite{ferriere01}. $X_{CO}$, on the other hand, is not well determined. Various models are used to tune $X_{CO}$ for obtaining the mass, column density and number density of ${\rm{H_{2}}}$. Most recently, a model of $X_{CO}$ has been developed to comply with $\gamma$-ray observations of \textit{Fermi}-LAT  \cite{ackermann12}.

\subsection{Neutral or atomic gas}

Neutral or atomic hydrogen, often denoted as HI, is another major component of interstellar gas. We can not directly detect HI in optical wavelengths. $\rm{HI}$ can be detected by observation of Lyman $\alpha$ (L$\alpha$) which  happens due to transition of electron between ground and first excited state of hydrogen atom at a wavelength of $1216 \AA$. The early L$\alpha$ survey was carried out by Savage and Jenkins \cite{savage1975} which was extended by Jenkins and Savage \cite{jenkins1974}. The outcome of such survey revealed the $\rm{HI}$ deficiency in the immediate vicinity of the Sun.  Later, it was known that the exact reason of such deficiency is related to the presence of the Sun inside the HI cavity commonly known as Local Bubble.   Subsequently, a deeper survey of L$\alpha$ by \textit{Copernicus satellite} \cite{bohlin1978} and \textit{International Ultraviolet Explorer (IUE)} \cite{shull1985} provided information of HI outside Local Bubble. But this diagnostic tool also suffers the same drawback of interstellar extinction. In this case, radio astronomy again helps us to overcome the difficulty. The remarkable breakthrough in the radio astronomical observation of HI happened with the detection of interstellar 21-cm line.  The 21-cm line arises due to the `hyperfine'  structure of the hydrogen atom. Briefly, we can say that the interaction of magnetic moment of electron and proton splits the electronic ground state in two close energy levels ; spin of electron is either parallel (upper energy level) or antiparallel (lower energy level) to the spin of proton. The transition between two energy levels corresponds to famous 21-cm line. This diagnostic tool is very useful to study HI in the ISM.  The vertical distribution of HI is modelled by assuming a Gaussian distribution \cite{ferriere01}.

\subsection{Ionized gas}
The information of ionized hydrogen gas, denoted as HII, was obtained from the radio signals coming from the pulsars and other (Galactic and extragalactic) compact objects. Cordes \textit{et al.} \cite{cordes1991} provided the space averaged of free electron density depending on the dispersion, distance and scattering measurements of pulsars. That calculation was later refined by Taylor and Cordes \cite{taylor1993}. In 2002, Cordes and Lazio \cite{cordes02} assembled all useful data and provide a non-axisymmetric  model of spatial distribution of interstellar free electrons, known as NE2001 model. To connect the density of free electrons with HII, we need to follow the conventional wisdom in which we assume that molecular and atomic media are fully neutral, while ionized media can be divided into two parts, namely warm ionized medium (WIM) and hot ionized medium (HIM) \cite{ferriere07}. Furthermore, we assume that helium is largely neutral in WIM and disregard the contribution of fully ionized helium in HIM \cite{ferriere07}. Considering such assumptions, we may identify density of HII as space averaged density of free electrons. Even if we consider the ionization of helium in HIM, the density of HII is still negligible compared to the contribution of WIM. Hence, we can say that HIM has less impact on the prediction of density of HII.

The interstellar dust is another component of ISM. But the discussion of such component is beyond the scope of the present work. In the next section, we study CR propagation with the \texttt{DRAGON} code.

\section{\label{sec:dragon}Modeling of cosmic ray propagation}

\noindent CRs, in our Galaxy, are generally believed to be accelerated by diffusive shock acceleration (DSA) mechanism at the shock regions of astrophysical objects \cite{bell78a,bell78b,blandford87} such as supernova remnants (SNRs). Such accelerated CRs are ultimately injected into the ISM and they then propagate through the stochastic magnetic field of ISM to reach at Earth. The observed energy spectrum, from sub GeV to multi-TeV, of CR is generally considered as a possible combination of  both the DSA and diffusive propagation processes occurring in our Galaxy. In the present work, we mainly study the propagation process of CRs from sources to Earth that can be modeled, for a given source distribution, gas density of ISM and injection spectra of primary CRs,  by solving transport equation \cite{gaggero12,strong07}. We, here, use a numerical code, called \texttt{DRAGON} \cite{bernardo10,evoli17}, to solve that transport equation. The numerical code, \texttt{DRAGON}, incorporates various physical processes such as scattering of CRs in the regular and turbulent Galactic magnetic field,  CR interaction in Galactic medium, radioactive decay of the nuclei and convection flow of CRs in the Galactic wind to obtain the solutions of the transport equation for the propagation of CRs in the Galaxy \cite{bernardo10,evoli17}. In the following part, we will choose suitable parameters  related to the geometry of our Galaxy, diffusion coefficient, injection spectra of primary CRs, models for CR source distribution, gas density and magnetic field structure of ISM.

\noindent We have used here the three dimensional model in the \texttt{DRAGON} code\footnote{The 3D version of the \texttt{DRAGON} code is available at \url{https://github.com/cosmicrays/DRAGON} for download.} for our present study. We have also considered a plain diffusion model (i.e., Alfven speed, $v_{A} = 0$) without any convection process during the transport of CRs in our Galaxy. 

 We have assumed CR sources follow a Ferriere-type \cite{ferriere01} spatial distribution. The primary particle populations such as protons and heavier nuclei, originated from the Ferriere-type source distribution, are injected into the ISM and their propagation are studied using \texttt{DRAGON} code. The injection spectra of proton and heavier nuclei are assumed to follow a power-law spectrum with a break such that the injection slope below the break is $\alpha_0$ and above it is $\alpha_1$. We have modeled the primary particles in the $0.1~{\rm{GeV}}-10~{\rm{TeV}}$ range with the assumed injection spectrum.
 
  We have assumed that the CRs propagate in a cylindrical region of radius ($R_{\rm{max}}$) with vertical boundary ($L$). We have further considered the spatially dependent diffusion coefficient that varies with particle rigidity ($\rho$) and the vertical height ($z$), above the Galactic plane, of the Galaxy. The form of such diffusion coefficient has been given below \cite{bernardo10,bernardo13}

  \begin{equation}
  D(\rho,z) = \beta^{\eta} D_{0}\Big( \frac{\rho}{\rho_{0}} \Big)^{\delta} {\rm{exp} }\Big( \frac{z}{z_{t}} \Big), 
\label{eq:diff}
\end{equation} 
where, $z_{t}$ and $\beta$ are  the halo height and particle speed, respectively. At low energy, uncertainties may arise due to propagation of CRs in the ISM \cite{bernardo11}. The power $\eta$ on $\beta$, in Eq.~(\ref{eq:diff}), accounts for such uncertainties at low energy. In Eq.~(\ref{eq:diff}), $D_{0}$ denotes  the normalization of diffusion coefficient and the reference rigidity is expressed as $\rho_{0}$. In the \texttt{DRAGON} code, we have selected the option `Pshirkov'  \cite{pshirkov11} for modeling the field structure of GMF. In this model, the GMF contains  three components namely disc, halo and turbulent, and the corresponding normalizations are denoted as $B_0^{\rm{disc}}$,  $B_0^{\rm{halo}}$ and $B_0^{\rm{turbulent}}$, respectively. It has been observed that disc component does not have significant effect on CR propagation whereas halo component has a role, albeit marginal \cite{bernardo13}. Recently, a different model of GMF with an extra X shaped component in the r-z (r is the radial distance from the Galactic center) plane  has been proposed \cite{jansson12}. But the new model does not significantly affect the result which is pointed out in Ref. \cite{bernardo13}. We, therefore, use the Pshirkov model for our analysis.

The turbulent component of GMF is another important component for the diffusion of CRs. On the basis of quasi-linear theory and the numerical simulations of particle propagation in turbulent magnetic fields \cite{marco07}, the z-dependence of the diffusion coefficient ($D(z)$) and the turbulent magnetic field ($B^{\rm{turbulent}}(z)$) is connected in the following way \cite{bernardo13} 

\begin{equation}
D(z)^{-1} \propto B^{\rm{turbulent}}(z) \propto {\rm{exp} }\Big(- z/z_{t} \Big)
\end{equation} 
where, $z_{t}$  can be treated as the effective scale-height of the diffusion region or the (magnetic) halo height. 
Theoretical modeling of the propagation of Galactic cosmic ray electrons and positrons to fit their observed flux, their
synchrotron emission and its angular distribution gave a relation between $B_0^{\rm{turbulent}}$ and $z_{t}$  \cite{bernardo13}. 
We note that in the above mentioned work  the transport equation in \texttt{DRAGON} code was solved by setting $L = 3z_{t}$ to avoid boundary effects.

 During the propagation, primary CRs interact with the gases (i.e., $\rm{H_{2}}$, HI and HII)  in the ISM and produce secondary CR nuclei. Such gas distributions are discussed below (see Sec.~\ref{sec:model}) in details. We also need to take into account the solar modulation effect which is dominant below $10$~GeV. To fit the observed CR spectra, we have modelled the solar modulation with a potential ($\phi$) such that the observed CR spectra can be modified by a factor \cite{usoskin05} 

\begin{equation}
\epsilon(E_{k}, Z,A,m_{Z}) = \frac{\Big( E_{k} + m_{Z}  \Big)^{2} - m_{Z}^{2}}{\Big( E_{k} + m_{Z} + \frac{Z |q|}{A}\phi \Big)^{2} - m_{Z}^{2} },  \label{eq:solarmod}
\end{equation}   
where, $q$ is the electronic charge unit, $m_{Z}$ is the mass of the nucleus having atomic number $Z$ and mass number $A$ , and $E_{k}$ is the kinetic energy of such nucleus.  

\begin{widetext}

\begin{table}[!h]
\caption{\label{tab:tabgdout} Various  cases of density profiles of hydrogen gas and the corresponding references for radial and vertical distributions are tabulated here. }
\begin{ruledtabular}
\begin{tabular}{ |p{0.8cm}|p{1 cm}|p{2.2cm}|l|p{2.2 cm}|l |}
Case  & Density profile & Radial distance from the Galactic center ($r$ in kpc) & Reference for radial distribution & Vertical distance above Galactic plane ($z$ in kpc)  &   Reference  for vertical distribution  \\
 \hline
 
  Case \#1 & $n_{{\rm{H_{2}}}}$ & For any $r$ & \cite{bronfman1988}, \cite{ackermann12} & For any $z$   & \cite{bronfman1988}, \cite{ackermann12} \\ \cline{2-6}
           & $n_{{\rm{HI}}}$  &  For $r < 16$~kpc  & \cite{gordon1976}, \cite{dickey1990} & For $r>10$~kpc and any $z$ & \cite{cox1986} \\\cline{5-6}
             &    &   &  & For $r<8$~kpc and any $z$ & \cite{dickey1990}   \\
\cline{2-6}
             & $n_{{\rm{HII}}}$ & For any r & \cite{cordes1991}  & For any $z$ & \cite{cordes1991} \\
 \hline
 Case \#2 &  $n_{{\rm{H_{2}}}}$ &  $r \lesssim3$~kpc  & \cite{ferriere07} & For any $z$   &  \cite{ferriere07} \\ \cline{3-6}
 &  & $r >3$~kpc  & \cite{ferriere1998} & For any $z$   &  \cite{ferriere1998} \\ 
 \cline{2-6}
 &  $n_{{\rm{HI}}}$ &  $r \lesssim 3$~kpc  & \cite{ferriere07} & For any $z$   &  \cite{ferriere07} \\ \cline{3-6}
 &  & $r >3$~kpc  & \cite{ferriere1998} & For any $z$   &  \cite{ferriere1998} \\ 
 \cline{2-6}
 &  $n_{{\rm{HII}}}$ &  $r \lesssim 3$~kpc  & \cite{ferriere07} & For any $z$   &  \cite{ferriere07} \\ \cline{3-6}
 &  & $r >3$~kpc  & \cite{ferriere1998} & For any $z$   &  \cite{ferriere1998} \\ 
 \hline
 
 Case \#3 &  $n_{{\rm{H_{2}}}}$ &  $r \lesssim 3$~kpc  & \cite{ferriere07} & For any $z$   &  \cite{ferriere07} \\ \cline{3-6}
 &  & $3~{\rm{kpc}}< r< 10.4$~kpc  & \cite{ferriere1998} & For any $z$   &  \cite{feldmann13} \\ 
 \cline{3-6}
  &  & $r \gtrsim 10.4$~kpc  & \cite{miller15} & For any $z$   &  \cite{feldmann13} \\ 
\cline{2-6}
        &  $n_{{\rm{HI}}}$ &  $r \lesssim 3$~kpc  & \cite{ferriere07} & For any $z$   &  \cite{ferriere07} \\ \cline{3-6}
 &  & $3~{\rm{kpc}}< r< 10.4$~kpc  & \cite{ferriere1998} & For any $z$   &  \cite{feldmann13} \\ 
 \cline{3-6}
  &  & $r \gtrsim 10.4$~kpc  & \cite{miller15} & For any $z$   &  \cite{feldmann13} \\ \cline{2-6}
   &  $n_{{\rm{HII}}}$ &  $r \lesssim 3$~kpc  & \cite{ferriere07} & For any $z$   &  \cite{ferriere07} \\ \cline{3-6}
 &  & $3~{\rm{kpc}}< r< 10.4$~kpc  & \cite{ferriere1998} & For any $z$   &  \cite{feldmann13} \\ 
 \cline{3-6}
  &  & $r \gtrsim 10.4$~kpc  & \cite{feldmann13} & For any $z$   &  \cite{feldmann13} \\ \cline{2-6}
 
\end{tabular}
\end{ruledtabular}
\end{table}

\end{widetext}

We have given a brief layout of our selected density profiles, mentioned as different cases, and references in Table~\ref{tab:tabgdout}. The information in Table~\ref{tab:tabgdout}  can act as a guideline to follow the detailed discussion of those density profiles  given in the next section.

\section{\label{sec:model}Different models of interstellar gas density profiles of hydrogen gas}

\noindent In this section, we will discuss three  different  cases of density profiles based on the observational results, hydrodynamical simulations and theoretical modelings. Each of the cases is implemented in the \texttt{DRAGON}  code to study the CR spectra.

\subsection{Case \#1}

 \begin{figure}[h!]
\subfigure{}
\includegraphics[width=0.45\textwidth,clip,angle=0]{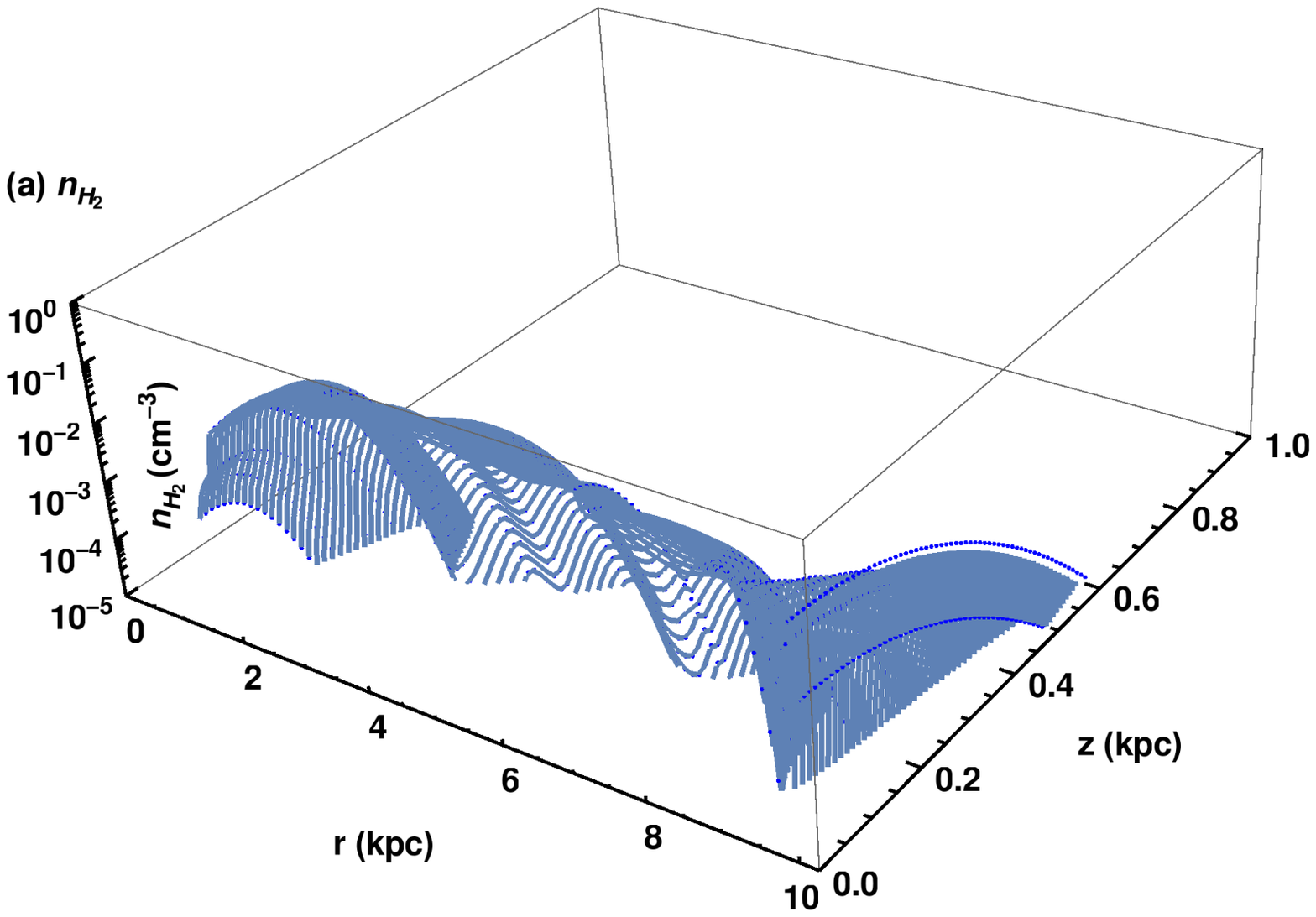}

\subfigure{}
\includegraphics[width=0.45\textwidth,clip,angle=0]{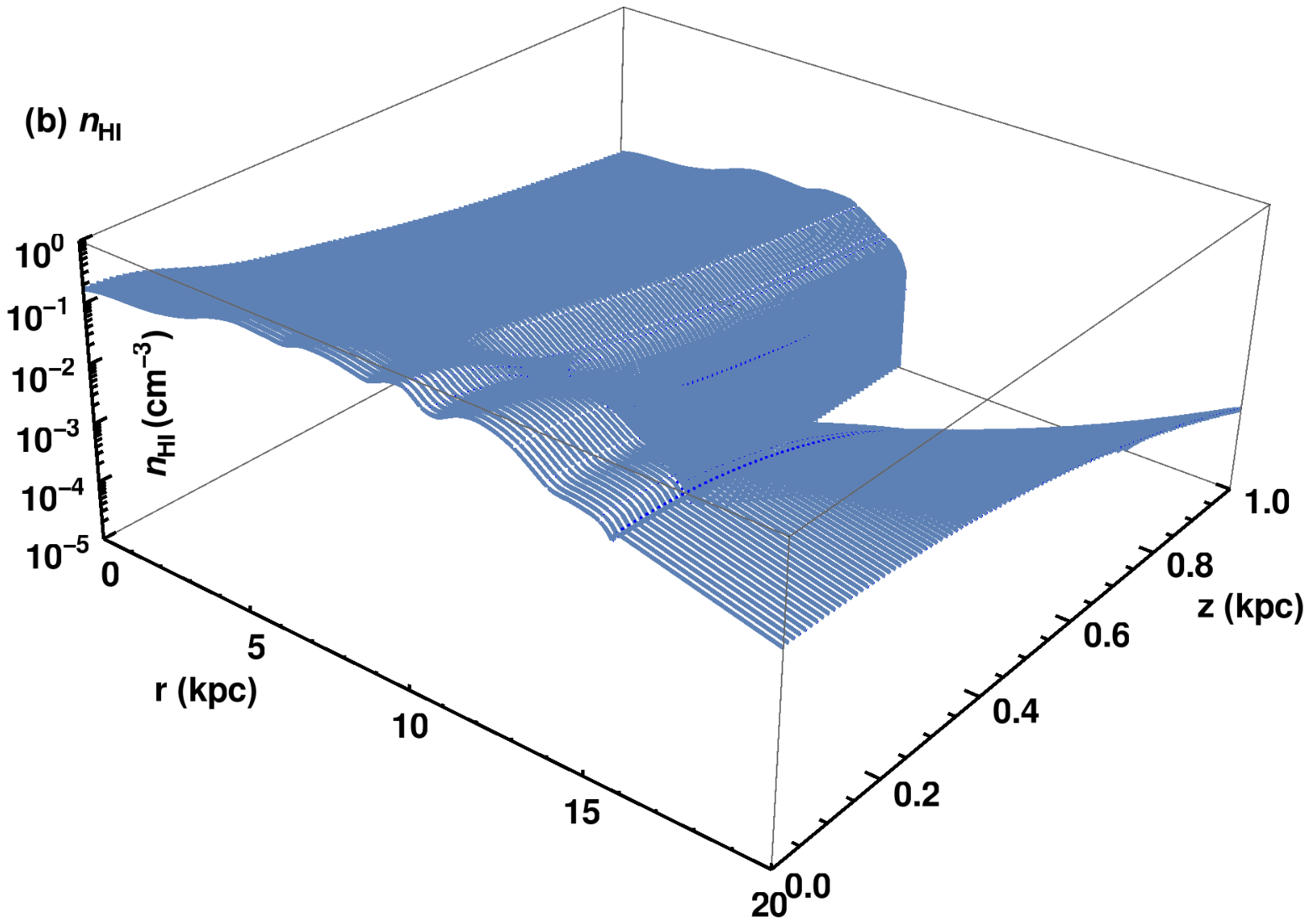}

\subfigure{}
\includegraphics[width=0.45\textwidth,clip,angle=0]{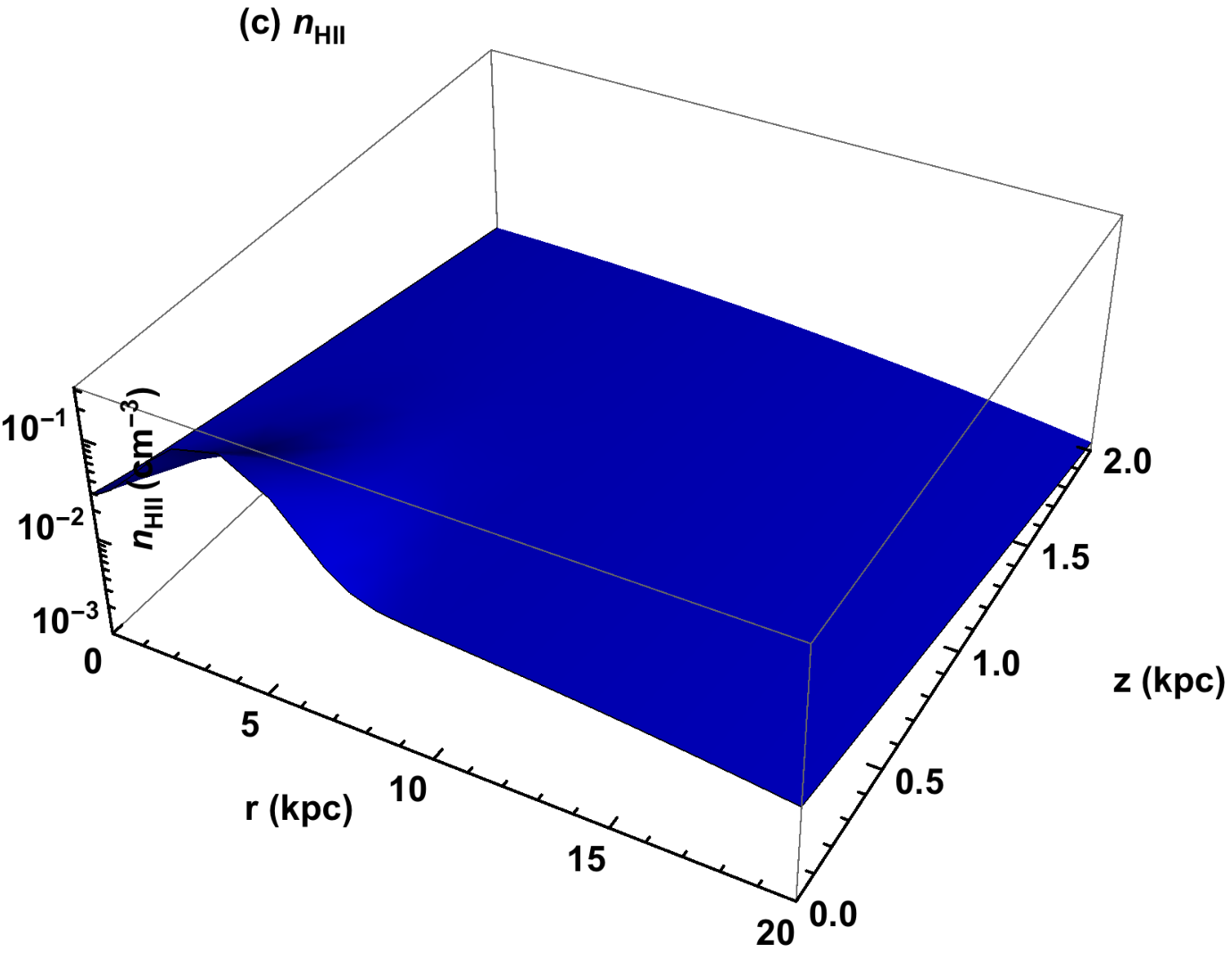}
 
 \caption{\label{fig:fig1} (Color online) The 3D plots of (a)  $n_{\rm{H_{2}}}$ \cite{bronfman1988, ackermann12}, (b)  $n_{\rm{HI}}$ \cite{gordon1976, cox1986, dickey1990}, and (c) $n_{\rm{HII}}$ \cite{cordes1991} with their radial and vertical distributions are shown. The density profiles are chosen in the \texttt{DRAGON} code by selecting the option `Galprop'. The $X_{\rm{CO}}$ model is selected as `galprop\textunderscore 2010' \cite{ackermann12}.}  
\end{figure}

\textbf{\Large{$n_{{\rm{H_{2}}}}$}:}  The density distribution of $\rm{H_{2}}$ gas in this case is given by 
 \begin{eqnarray}
 n_{\rm{H_{2}}} (r, z) = \epsilon_{0} (r)~ X_{{\rm{CO}}}~{\rm{exp}}\Bigg[ -\frac{{\rm{ln}} 2 \big(z - z_{o}(r)\big)^{2}}{z_{h}(r)^{2}}\Bigg]  \nonumber  \\
 \times (1.0/3.08 \times 10^{21})~ \rm{cm^{-3} },
\label{eq:gd01}
\end{eqnarray}  
where, $r$ is the radial distance from the Galactic center and $z$ is the height above the Galactic plane. In Eq.~(\ref{eq:gd01}), $\epsilon_{0} (r)$, $z_{0}(r)$ and $z_{h}(r)$ are the CO volume emissivity (in unit of $\rm{K~km~s^{-1}}$),  the scale height and width, respectively. Here, $X_{{\rm{CO}}}$ denotes the conversion factor from CO to  $\rm{H_{2}}$ (in unit of $\rm{cm^{-2}~K^{-1}km^{-1} s}$). The values of $\epsilon_{0} (r)$, $z_{0}(r)$ and $z_{h}(r)$ are taken from Cols. 4, 7, 10 of Table~3 of Ref.~\cite{bronfman1988}. The data in Table~3 of Ref.\cite{bronfman1988} is based on the CO survey of Southern and Northern Milky Way. The Southern survey of CO was done by using the Columbia  1.2 m Millimeter-Wave Telescope situated at Cerro Tololo, Chile.The Northern CO Survey was completed with the Columbia Telescope in New York City. $X_{{\rm{CO}}}$  is a very uncertain parameter. It is chosen with an option  `galprop\textunderscore 2010' in the code  such that it is compatible with the Fermi-LAT $\gamma$-ray observations \cite{ackermann12}. The 3D plot of $n_{\rm{H_{2}}} (r, z)$ is shown in Fig.1(a).

\textbf{\Large{$n_{{\rm{HI}}}$}:} The density profile  of HI gas in the \texttt{DRAGON} code can be expressed as 
 \begin{equation}
 n_{{\rm{HI}}}(r,z) = Y(r)f(z),
 \label{eq:gd02}
\end{equation}  
 where, $Y(r)$ is the renormalized radial distribution and $f(z)$ denotes the vertical distribution. The  $n_{{\rm{HI}}}(r,z)$ (for $r < 16$~kpc), here, is based on the report of  HI relative distribution as provided in the Table~1 of Ref.~\cite{gordon1976}. However, the data of Ref.~\cite{gordon1976} is renormalized to satisfy the result of Ref.~\cite{dickey1990} based on the data of Lyman-$\alpha$ and 21-cm line.  Finally, the vertical distribution, for $r > 10$~kpc, is obtained on the basis of theoretical model of infrared (IR) or sub mm emission from the Galactic disk as provided in Ref.~\cite{cox1986}.  Lyman-$\alpha$ and 21-cm line data in Ref.~\cite{dickey1990} provides vertical distribution for $r < 8$~kpc and a proper interpolation is done to get the vertical distribution in between 8 and 10 kpc. For $r > 16$~kpc, an extrapolation has been done by considering an exponential tail of the data for $r < 16$~kpc with a scale height of $3$~kpc. The 3D plot of $n_{{\rm{HI}}}(r,z)$ is shown in Fig.1(b).

\textbf{\Large{$n_{{\rm{HII}}}$}:}  The expression for density profile of HII is considered in the \texttt{DRAGON} code as \cite{cordes1991}
  
  \begin{eqnarray}
 n_{{\rm{HII}}}(r,z) =&& (0.025~ \rm{cm^{-3} }) \nonumber \\
 && \times {\rm{exp}}\Bigg[-\bigg(\frac{|z|}{1.0~{\rm{kpc}}}\bigg)\Bigg] {\rm{exp}}\Bigg[-\bigg(\frac{r}{20.0~{\rm{kpc}}}\bigg)^{2}\Bigg] \nonumber \\
   &&+  (0.20~ \rm{cm^{-3} })  {\rm{exp}}\Bigg[-\bigg(\frac{|z|}{0.15~{\rm{kpc}}}\bigg)\Bigg] \nonumber \\
   && \times {\rm{exp}}\Bigg[-\bigg(\frac{r- 4.0~{\rm{kpc}}}{2.0~{\rm{kpc}}}\bigg)^{2}\Bigg]. 
\label{eq:gd03}  
\end{eqnarray}

The above density profile is obtained from dispersion, distance and radio wave scattering measurements of pulsars. In 1991, Cordes \textit{et al.} \cite{cordes1991} analyzed two distinct pulsar data sets. The data sets contain either measurements of dispersion and distance or measurements of dispersion and scattering. The analyzed data sets are then fitted with a two-component axisymmetric model of free electron density to obtain the above density profile. The 3D plot of $ n_{{\rm{HII}}}(r,z) $ is shown in Fig.1(c).

\subsection{Case \#2}

For our study, we have considered two regions; one is $r \lesssim 3$~kpc, known as Galactic bulge (GB) and other one is $r >3$~kpc.

\subsubsection{For $r \lesssim 3$~kpc  }

In this region, the density profiles are based on the observational results summarized by Morris and Serabyn \cite{morris1996} and Mezger \textit{et al.} \cite{mezger1996}. The results are based on the observational studies on CO line emission, 21-cm emission and absorption lines, thermal emission from dusts. The observational results are also complemented with theoretical predictions obtained from the gas dynamical models near the Galactic center (GC). Later, Ferriere \textit{et al.} \cite{ferriere07} provides a theoretical model based on such observation results and gas dynamical models. We have considered the model of  Ferriere \textit{et al.} \cite{ferriere07} for our present work.

\begin{figure}[h!]
\subfigure{}
\includegraphics[width=0.45\textwidth,clip,angle=0]{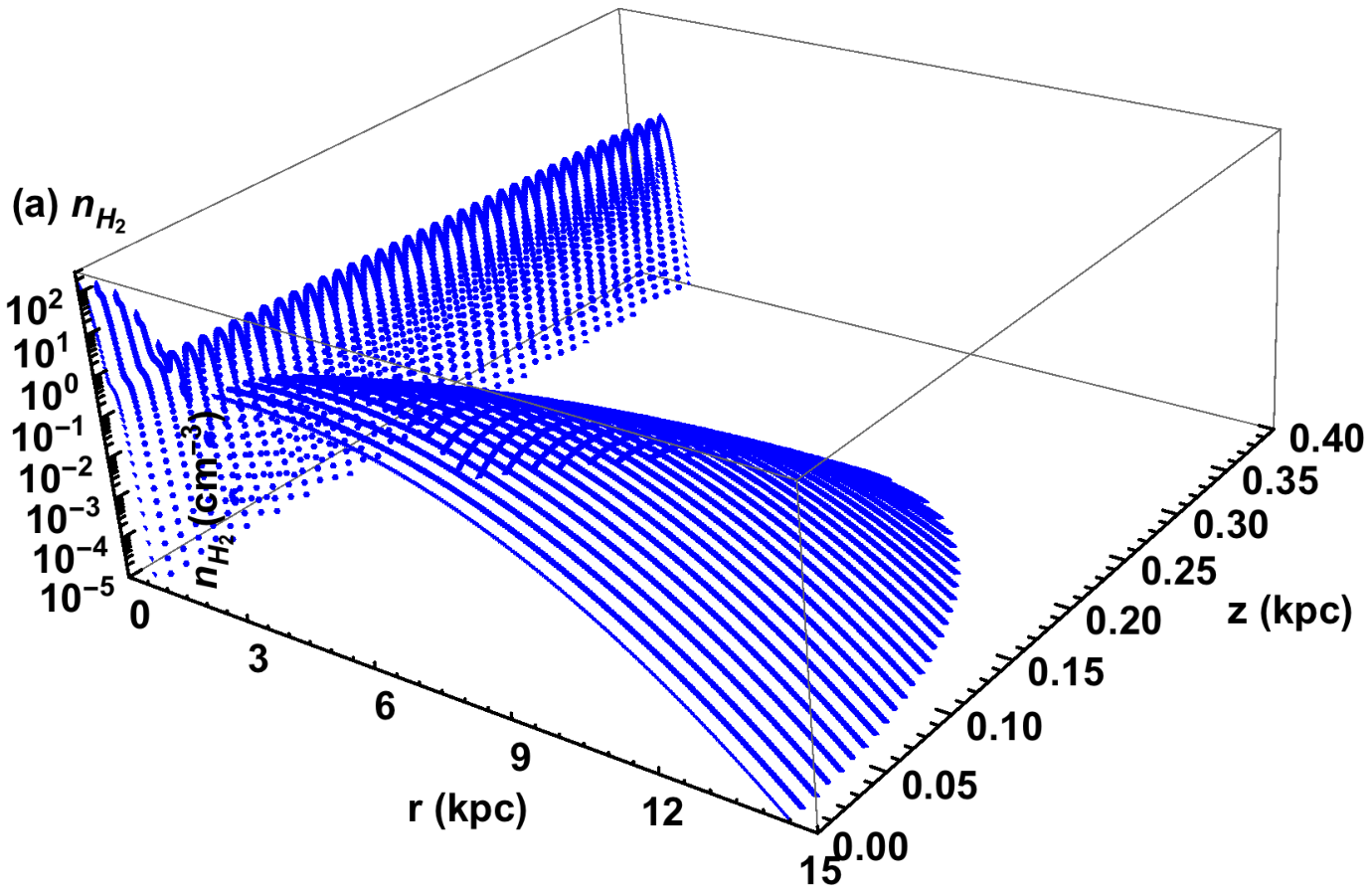}

\subfigure{}
\includegraphics[width=0.45\textwidth,clip,angle=0]{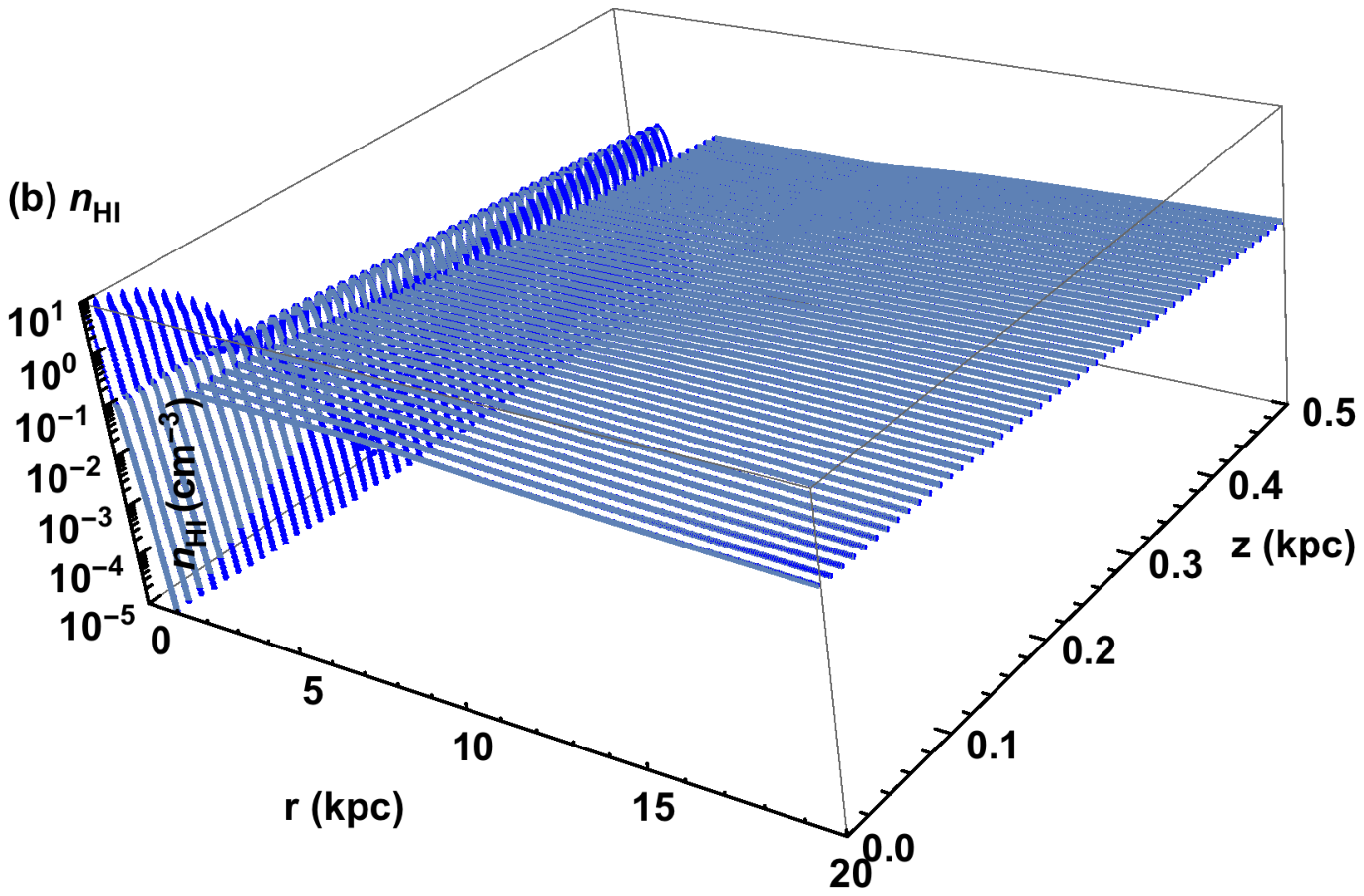}

\subfigure{}
\includegraphics[width=0.45\textwidth,clip,angle=0]{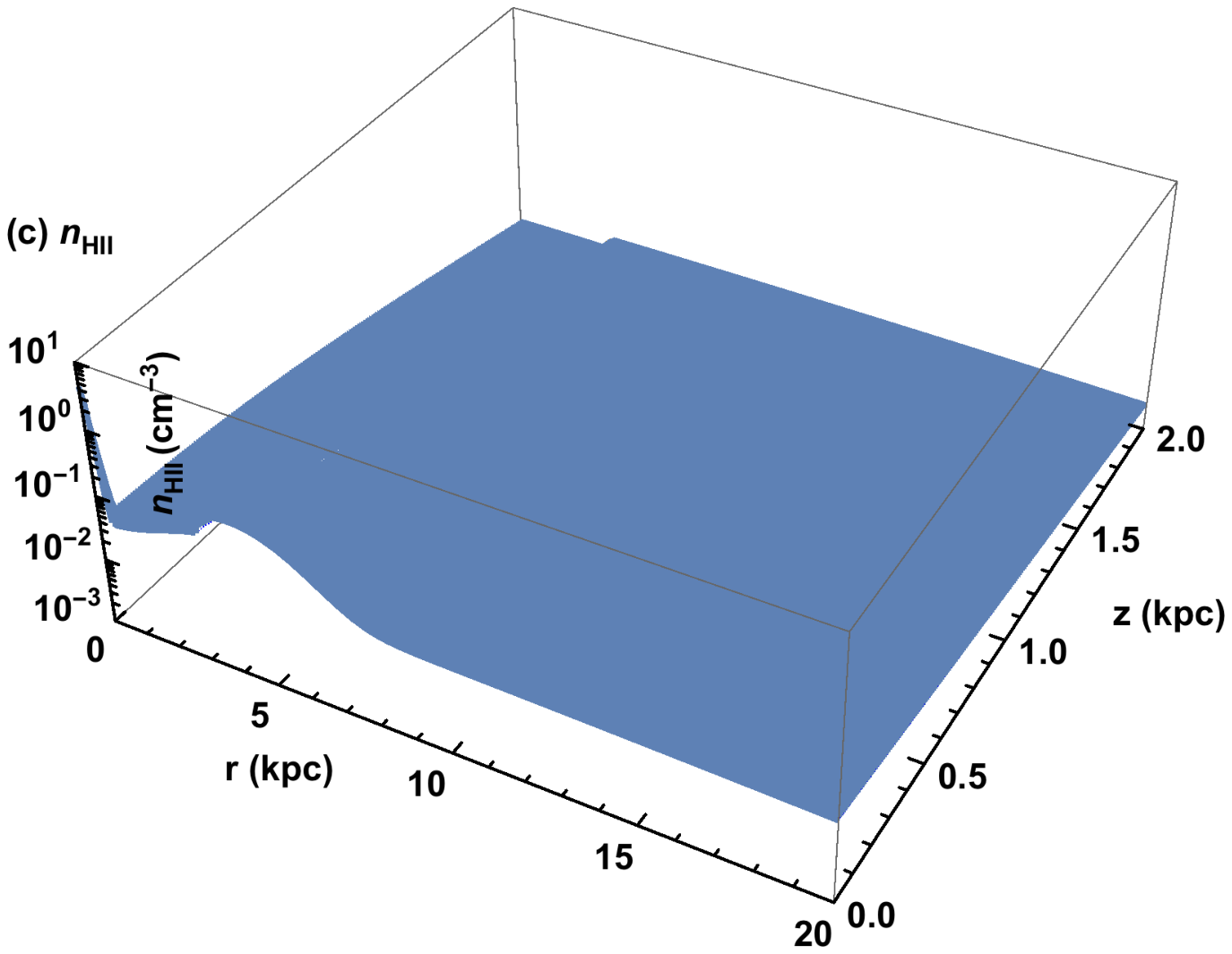}

\caption{\label{fig:fig3} (Color online) The 3D plots of (a)  $n_{\rm{H_{2}}}$ \cite{ferriere07,ferriere1998}, (b)  $n_{\rm{HI}}$ \cite{ferriere07,ferriere1998}, and (c) $n_{\rm{HII}}$ \cite{ferriere07, ferriere1998} with their radial and vertical dependence are shown. The plots correspond to density profiles mentioned in Case \#2. In case of  $n_{\rm{H_{2}}}$, $n_{\rm{HI}}$, a gap, in the range 1-3 kpc, has been occurred due to Galactic bar effect. }
\end{figure}

\textbf{\Large{$n_{{\rm{H_{2}}}}$}:}  In this region, the density profile is a combined contribution of central molecular zone (CMZ) and the GB disk. The CMZ is an asymmetric layer of molecular gas extending up to $r \sim 200$~pc and $r \sim 150$~pc in the positive and negative longitudes, respectively \cite{ferriere01}. The CMZ also contains a ring like feature having a mean radius of $r\sim 180$~pc which is termed as 180-pc molecular ring.

 The density profile, in this case, is based on the work of Sawada \textit{et al.} \cite{sawada04}. This work provides a face on map solely based on the observations. The distribution of molecular gas has been obtained from the quantitative comparison of 2.6 mm CO emission line with the 18 cm OH absorption line. Depending on the study of Sawada \textit{et al.} \cite{sawada04}, we can visualize the  CMZ projected onto the Galactic plane has an elliptic shape ($500~{\rm{pc}} \times 200~{\rm{pc}}$) making an angle of $70^{\circ}$ with the line of sight toward positive longitudes \cite{ferriere07}. The center of the ellipse is at $(x_{c}, y_{c})\simeq (-50 ~{\rm{pc}}, 50~{\rm{pc}})$ \cite{ferriere07}. The relation between CMZ coordinates ($X, Y$) and the Galactic coordinates ($x, y$) are related by the following equations \cite{ferriere07}
 
 \begin{eqnarray}
 X = (x-x_{c})~{\rm{cos} } \theta_{c} + (y-y_{c})~{\rm{sin }} \theta_{c} \\
Y =- (x-x_{c})~{\rm{sin} } \theta_{c} + (y-y_{c})~{\rm{cos }} \theta_{c},
\end{eqnarray}
where, $x_{c} =-50~{\rm{pc}},  y_{c} = 50~{\rm{pc}}$ and $\theta_{c} = 70^{\circ}$. We, however, do not get any information of vertical distribution from the work of Sawada \textit{et al.} \cite{sawada04}. For the vertical distribution, we have followed the work of  Burton and Liszt \cite{burton1992} and Gaussian distribution is assumed as a possible form. The normalization of the density profile in the region is obtained by considering the $\rm{H_{2}}$ mass to be $1.9 \times 10^{7} M_{\odot}$ (assuming $X_{{\rm{CO}}} = 0.5 \times 10^{20}~\rm{cm^{-2}~K^{-1}km^{-1} s}$). The density distribution in the CMZ can be expressed as \cite{ferriere07} 

\begin{eqnarray}
n_{{\rm{H_{2}}}}^{\rm{CMZ}} =&& (150.0~{\rm{cm^{-3}}}) \nonumber \\
&& \times {\rm{exp}}\Bigg[- \Bigg(\frac{\sqrt{X^2 + (2.5Y)^2}-0.125~{\rm{kpc}}} {0.137~{\rm{kpc}}}\Bigg)^{4}\Bigg] \nonumber \\
&& \times {\rm{exp}}\Bigg[- \bigg(\frac{z}{0.018~{\rm{kpc}}}\bigg)^2\Bigg].
\end{eqnarray}

The GB disk (region beyond CMZ), based on the work of Liszt and Burton \cite{listz1980}, has also an elliptical shape with a hole at the central region. The GB disk is tilted out of the Galactic plane by $\alpha = 13.5^{\circ}$ (rotated counterclockwise about the $x$-axis, where, $(x,y,z) \rightarrow (x,y^{\prime}, z^{\prime})$ and $(x,y,z)$ is the Galactic coordinate), inclined by $\beta = 20 ^{\circ}$ (rotation about the $y^{\prime}$ axis, whereby $(x,y^{\prime}, z^{\prime} \rightarrow (x^{\prime \prime},y^{\prime}, z^{\prime \prime})$, and the major axis of the GB disk forms an angle $\theta_{d} = 48.5 ^{\circ}$ to the $x^{\prime \prime}$ axis. The coordinates of GB disk ($\cal{X, Y, Z}$) are related to the coordinates of our Galaxy ($x,y,z$) in the following way \cite{ferriere07}

\begin{eqnarray}
{\cal{X}} =&&  x ~{\rm{cos} } \beta~{\rm{cos} } \theta_{d} \nonumber \\
 &&-y~ ( {\rm{sin} } \alpha ~{\rm{sin} } \beta ~{\rm{cos} } \theta_{d}  - {\rm{cos} } \alpha ~{\rm{sin} } \theta_{d} ) \nonumber \\
 && -z~({\rm{cos} } \alpha ~{\rm{sin} } \beta ~{\rm{cos} } \theta_{d}  + {\rm{sin} } \alpha~{\rm{sin} } \theta_{d})   \\
 {\cal{Y}} =&&  -x ~{\rm{cos} } \beta ~{\rm{sin} } \theta_{d} \nonumber \\
 &&+y~ ( {\rm{sin} } \alpha ~{\rm{sin} } \beta ~{\rm{sin} } \theta_{d}  + {\rm{cos} } \alpha ~{\rm{cos} } \theta_{d} ) \nonumber \\
 && +z~({\rm{cos} } \alpha ~{\rm{sin} } \beta~{\rm{sin} } \theta_{d}  - {\rm{sin} } \alpha ~{\rm{cos} } \theta_{d})   \\
 {\cal{Z}} =&&  x ~{\rm{sin} } \beta\nonumber \\
 &&+y~  {\rm{sin} } \alpha ~{\rm{cos} } \beta \nonumber \\
 && +z~{\rm{cos} } \alpha ~{\rm{cos} } \beta. 
\end{eqnarray}

The normalization of the density profile in the region is obtained by considering the $\rm{H_{2}}$ mass to be $3.4 \times 10^{7} M_{\odot}$. The density distribution in the holed GB disk can be expressed as \cite{ferriere07} 
  
\begin{eqnarray}
n_{{\rm{H_{2}}}}^{\rm{disk}} =&& (4.8~{\rm{cm^{-3}}}) \nonumber \\
&& \times {\rm{exp}}\Bigg[- \Bigg(\frac{\sqrt{{\cal{X}}^2 + (3.1{\cal{Y}})^2}-1.2~{\rm{kpc}}} {0.438~{\rm{kpc}}}\Bigg)^{4}\Bigg] \nonumber \\
&& \times {\rm{exp}}\Bigg[- \bigg(\frac{{\cal{Z}}}{0.042~{\rm{kpc}}}\bigg)^2\Bigg].
\end{eqnarray}

The total density distribution of $\rm{H_{2}}$ can be written as 

\begin{equation}
n_{{\rm{H_{2}}}} (r,z)= n_{{\rm{H_{2}}}}^{\rm{CMZ}}  + n_{{\rm{H_{2}}}}^{\rm{disk}}.     
\label{eq:h2fer07}
\end{equation}
\\

\textbf{\Large{$n_{{\rm{HI}}}$}:}  Similar to $\rm{H_{2}}$,  the density distribution of $\rm{HI}$ comes from the combined contribution of CMZ and holed GB disk.

From different surveys of CMZ, we may conclude that mass of $\rm{HI}$ is $8.8\%$ of  the mass of $\rm{H_{2}}$ in that region. The space-averaged density of ${\rm{HI}}$ can be written as \cite{ferriere07}

\begin{eqnarray}
n_{{\rm{HI}}}^{\rm{CMZ}} =&& (8.8~{\rm{cm^{-3}}}) \nonumber \\
&&\times {\rm{exp}}\Bigg[- \Bigg(\frac{\sqrt{X^2 + (2.5Y)^2}-0.125~{\rm{kpc}}} {0.137~{\rm{kpc}}}\Bigg)^{4}\Bigg] \nonumber \\
&& \times {\rm{exp}}\Bigg[- \bigg(\frac{z}{0.054~{\rm{kpc}}}\bigg)^2\Bigg].
\end{eqnarray}
 
 Similarly, the space-averaged density of $\rm{HI}$ in the holed GB disk is the following \cite{ferriere07}

\begin{eqnarray}
n_{{\rm{HI}}}^{\rm{disk}} =&& (0.34~{\rm{cm^{-3}}}) \nonumber \\
&& \times {\rm{exp}}\Bigg[- \Bigg(\frac{\sqrt{{\cal{X}}^2 + (3.1{\cal{Y}})^2}-1.2~{\rm{kpc}}} {0.438~{\rm{kpc}}}\Bigg)^{4}\Bigg] \nonumber \\
&& \times {\rm{exp}}\Bigg[- \bigg(\frac{{\cal{Z}}}{0.120~{\rm{kpc}}}\bigg)^2\Bigg].
\end{eqnarray}

So, the total density distribution of ${\rm{HI}}$ can be written as 

\begin{equation}
n_{{\rm{HI}}} (r,z)= n_{{\rm{HI}}}^{\rm{CMZ}}  + n_{{\rm{HI}}}^{\rm{disk}}.     
\label{eq:hifer07}
\end{equation}
\\

\textbf{\Large{$n_{{\rm{HII}}}$}:} The density distribution of ionized component is based on the NE2001 model. The model is a non-axisymmetric of the spatial distribution of free electrons in our Galaxy developed on the basis of the data of dispersion, scattering and distance measurements of pulsars available at the end of 2001. 

In our case, we have only considered the contribution of WIM as it contributes $83\%$ of the total mass of HII. We have assumed that helium is completely neutral and hydrogen gas is completely ionized. The space-averaged density of HII is given as \cite{ferriere07} 

\begin{eqnarray}
n_{{\rm{HII}}}(r,z) =&&  (8.0~{\rm{cm^{-3}}}) \nonumber \\
&& \times \Bigg\{ {\rm{exp}} \Bigg[  -  \frac{x^{2} + (y - y_{3})^{2}}{L_{3}^{2}}  \Bigg] \nonumber \\
&&\times {\rm{exp}} \Bigg[- \frac{(|z|-{z_{3})^{2}}}{H_{3}^{2}}   \Bigg] \nonumber \\
&&+ 0.009 \times {\rm{exp}}\Bigg[ - \Bigg(\frac{r - L_{2}}{L_{2}/2}\Bigg)^{2}  \Bigg]  {\rm{sech^{2}}} \bigg(\frac{|z|}{H_{2}}\bigg) \nonumber \\
&&+ 0.005\Bigg[ {\rm{cos}} \Bigg( \pi \frac{r}{2L_{1}}  \Bigg) u(L_{1} - r)  \Bigg]  \nonumber \\
&&  \times {\rm{sech^{2}}} \bigg(\frac{|z|}{H_{1}}\bigg)     \Bigg\},
\end{eqnarray}  

where, $u$ is the unit step function, $y_{3} =-10~{\rm{pc}}$, $z_{3} =-20~{\rm{pc}}$, $L_{3} =145~{\rm{pc}}$, $H_{3} =26~{\rm{pc}}$, $L_{2} =3.7~{\rm{kpc}}$, $H_{2} =140~{\rm{pc}}$,$L_{1} =17~{\rm{kpc}}$ and  $H_{1} =950~{\rm{pc}}$.

\subsubsection{For $r > 3$~kpc }

\textbf{\Large{$n_{{\rm{H_{2}}}}$}:}  In this region, the radial density profile of $\rm{H_{2}}$ exhibits a weaker peak (a stronger peak exists at the Galactic center) at a Galactocentric distance $\sim 4.5$~kpc. Along with the radial distribution, a Gaussian distribution is also adopted for expressing the vertical distribution of  $\rm{H_{2}}$.  The $X_{{\rm{CO}}}$ and the full width at the half maximum (FWHM) of the Gaussian distribution are obtained by analyzing the data of Massachusetts-Stony Brook Galactic Plane CO survey \cite{clemens1988}. Such survey was based on the    
2.6 mm CO emission line. The space-averaged number density of $n_{{\rm{H_{2}}}}$ can be expressed as \cite{ferriere1998}

\begin{widetext}
\begin{eqnarray}
n_{{\rm{H_{2}}}} (r,z)= && (0.5 \times 0.58~{\rm{cm^{-3}}}) \times \Bigg(  \frac{r}{8.5~{\rm{kpc}}}  \Bigg)^{-0.58} \nonumber \\
 && \times {\rm{exp}}\Bigg[  - \frac{(r - 4.5~{\rm{kpc}})^{2}  -  (4.0~{\rm{kpc}})^{2}   } {(2.9~{\rm{kpc}})^{2}}   \Bigg]   \nonumber \\
  &&  \times  {\rm{exp}} \Bigg[ - \bigg(  \frac{z}{0.081~{\rm{kpc}}}   \bigg)^{2} \bigg(  \frac{r}{8.5~{\rm{kpc}}}   \bigg)^{-1.16}     \Bigg].
 \label{eq:h2fer98}
\end{eqnarray}
\end{widetext}

\textbf{\Large{$n_{{\rm{HI}}}$} :} The space-averaged density profile of HI, developed on the basis of 21 cm emission and absorption line data, is given by \cite{ferriere1998}

\begin{widetext}
 \begin{eqnarray}
n_{{\rm{HI}}}(r,z) = &&\frac{(0.340~{\rm{cm^{-3}}} )}{(\alpha_{h} (r))^{2}}  \times \Bigg\{   0.859~{\rm{exp}}\Bigg[ - \Bigg(\frac{z}{(0.127~{\rm{kpc}})\alpha_{h}(r)}   \Bigg)^{2} \Bigg] \nonumber \\
&&  +   0.047~{\rm{exp}}\Bigg[ - \Bigg(\frac{z}{(0.318~{\rm{kpc}})\alpha_{h}(r)}   \Bigg)^{2} \Bigg] + 0.094~{\rm{exp}}\Bigg[ - \Bigg(\frac{|z|}{(0.403~{\rm{kpc}})\alpha_{h}(r)}   \Bigg) \Bigg]      \Bigg\}  \nonumber \\
 && + \frac{(0.226~{\rm{cm^{-3}}} )}{(\alpha_{h} (r))}  \nonumber \\
&&  \times \Bigg\{\Bigg[  1.745 - \frac{1.289}{\alpha_{h} (r))} \Bigg] \times  {\rm{exp}}\Bigg[ - \Bigg(\frac{z}{(0.127~{\rm{kpc}})\alpha_{h}(r)}   \Bigg)^{2} \Bigg]                       \nonumber \\
 && +  \Bigg[ 0.473 - \frac{0.070}{\alpha_{h} (r))} \Bigg]  \times {\rm{exp}}\Bigg[ - \Bigg(\frac{z}{(0.318~{\rm{kpc}})\alpha_{h}(r)}   \Bigg)^{2} \Bigg]                       \nonumber \\ 
 && +  \Bigg[  0.283 - \frac{0.142}{\alpha_{h} (r))} \Bigg] \times  {\rm{exp}}\Bigg[ - \Bigg(\frac{|z|}{(0.403~{\rm{kpc}})\alpha_{h}(r)}   \Bigg) \Bigg] \Bigg\}.  
\label{eq:hifer98}
\end{eqnarray}
\end{widetext}
 where,
\begin{equation}
\begin{rcases}
 \alpha_{h} (r) &= 1.0,~ {\rm{For,~ r \leq 8.5~kpc}}  \\
                                      &= \frac{r}{8.5 ~{\rm{kpc}}}, ~ {\rm{For,~ r > 8.5~kpc}}.\\
\end{rcases}
\end{equation}

\textbf{\Large{$n_{{\rm{HII}}}$}:} In this case, we have considered only the contribution of WIM. From the dispersion, scattering and distance measurements of pulsars we can express the space-averaged density distribution of $\rm{HII}$ as \cite{ferriere1998}

\begin{eqnarray}
 n_{{\rm{HII}}}(r,z) =&& (0.0237~{\rm{cm^{-3}}} )  {\rm{exp}}\Bigg[-\frac{r^{2} - (8.5~{\rm{kpc}})^{2} }{(37.0~{\rm{kpc}})^{2}}\Bigg] \nonumber \\
 && \times  {\rm{exp}}\Bigg( - \frac{|z|}{1.0~{\rm{kpc}}}   \Bigg) \nonumber \\
   &&+ (0.0013~{\rm{cm^{-3}}} )  \nonumber \\
  && \times  {\rm{exp}}\Bigg[-\frac{(r- 4.0~{\rm{kpc}})^{2} - (4.5~{\rm{kpc}})^{2}    }{(2.0~{\rm{kpc}})^{2}} \Bigg] \nonumber \\
&& \times   {\rm{exp}}\Bigg( - \frac{|z|}{0.150~{\rm{kpc}}}   \Bigg).
\label{eq:hiifer98}
\end{eqnarray}

In Figs.~2(a,b,c), we have shown the density profiles of molecular, atomic and ionized hydrogen gas. In the $r \lesssim 3$~kpc region, we have considered a co-ordinate transformation, i.e. from Cartesian ($x,y,z$) to cylindrical ($r, \theta, z$) and taken the azimuthal average of Eqs.~(14), (17) and (18) to obtain $n_{{\rm{H_{2}}}}(r.z)$, $n_{{\rm{HI}}}(r.z)$, and $n_{{\rm{HII}}}(r.z)$. In Figs.~2(a,b), a gap in the range 1-3 kpc has been observed corresponding to $\rm{H_{2}}$ and $\rm{HI}$. Actually, the amount of molecular and atomic gas is indeed observed to be very low in the radial range ~ 1 - 3 kpc. The reason for this gap in the gas distributions can be explained in terms of the effects of the Galactic bar. It turns out that particle orbits tend to be unstable between the bar's corotation radius and its outer Lindblad resonance (OLR) \cite{ferriere07}. Interstellar gas tends to be expelled from the unstable region, thereby creating a gap between corotation and the OLR \cite{ferriere07}. Gas expelled inwards accumulates inside corotation, where it gives rise to the tilted disk of the GB, and deeper in (i.e., inside the bar's inner Lindblad resonance) to CMZ. Gas expelled outwards accumulates outside the OLR, where it gives rise to the molecular ring of the Galactic disk \cite{ferriere07}. To plot the density profiles, proper interpolation, whenever needed, is done between the data of different regions.

\subsection{Case \#3}

We, here, consider another case. For such purpose, we have divided our region of interest in two regions (similar to Case \#2); one is $r\lesssim 3$~kpc and other one is $r > 3$~kpc.

\begin{figure}[h!]
\subfigure{}
\includegraphics[width=0.45\textwidth,clip,angle=0]{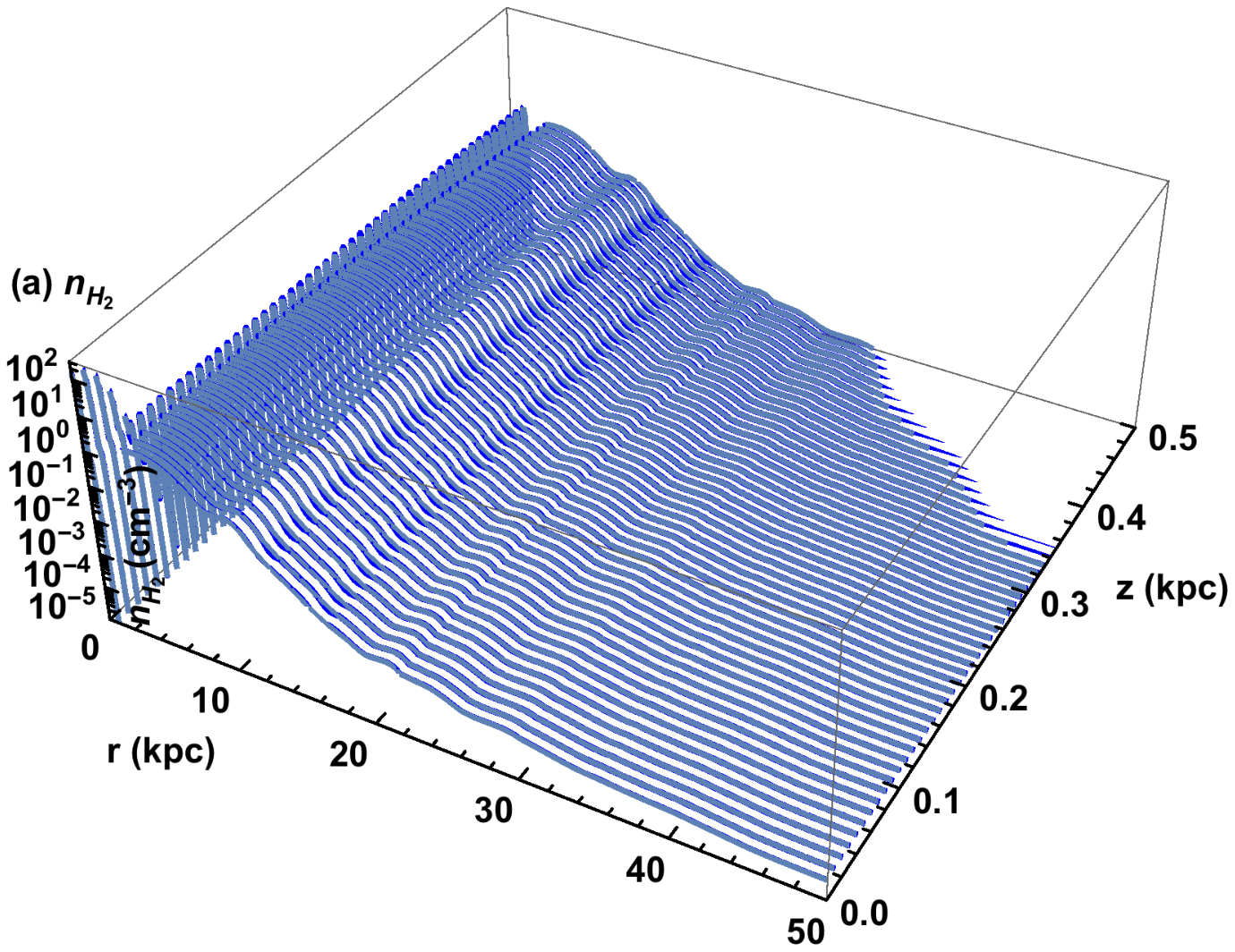}

\subfigure{}
\includegraphics[width=0.45\textwidth,clip,angle=0]{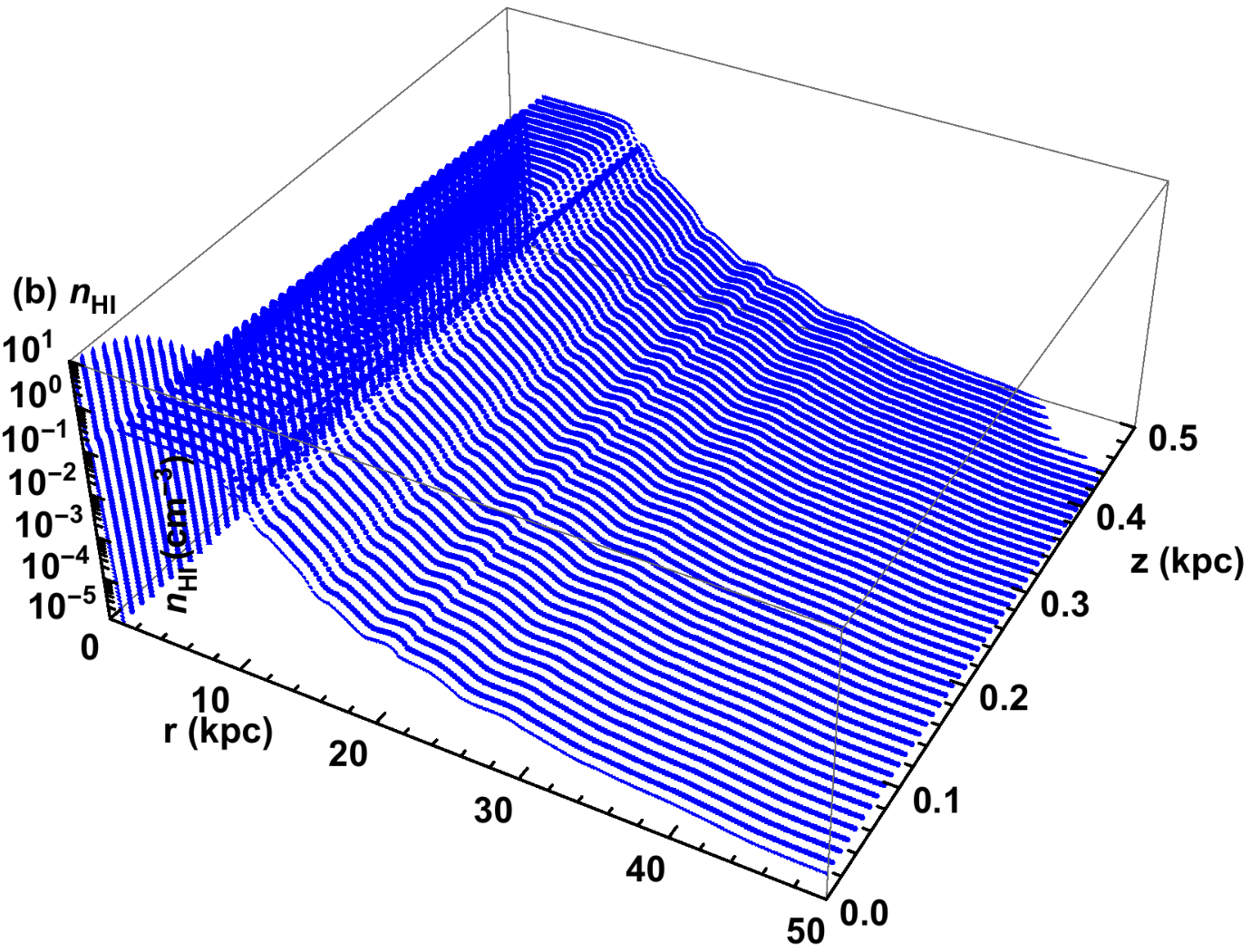}

\subfigure{}
\includegraphics[width=0.45\textwidth,clip,angle=0]{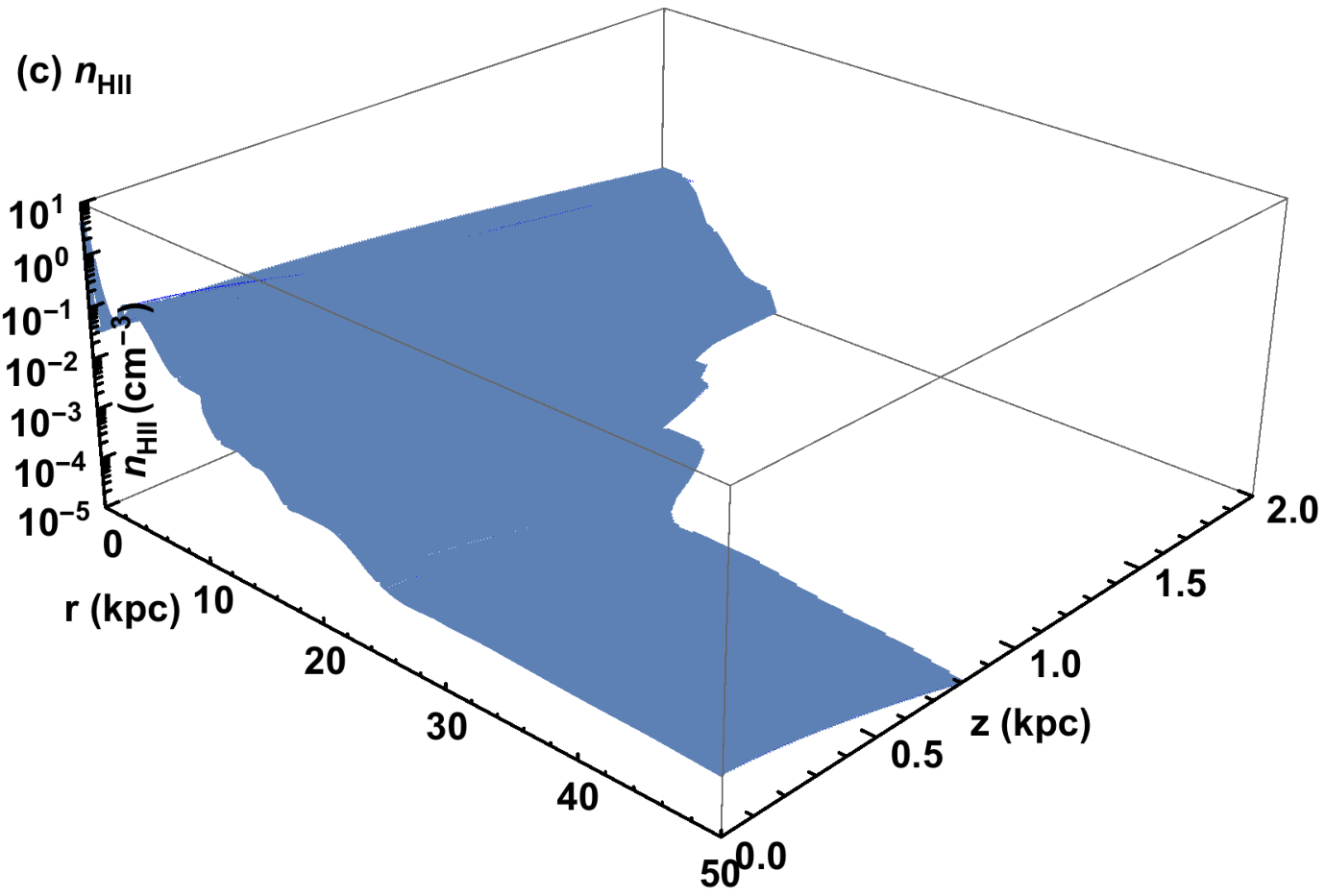}

\caption{\label{fig:fig4} (Color online) The 3D plots of (a)  $n_{\rm{H_{2}}}$ \cite{ferriere07,ferriere1998, feldmann13,miller15}, (b)  $n_{\rm{HI}}$ \cite{ferriere07,ferriere1998,feldmann13,miller15}, and (c) $n_{\rm{HII}}$ \cite{ferriere07, ferriere1998,feldmann13,miller15} with their radial and vertical dependence are shown. The plots correspond to density profiles mentioned in Case \#3. In case of  $n_{\rm{H_{2}}}$, $n_{\rm{HI}}$, a gap, in the range 1-3 kpc, has been occurred due to Galactic bar effect.  The  radial distributions, i.e., $r\gtrsim10.4$~kpc and vertical distributions ($> 3$~kpc) are modified, with respect to Case \#2, to comply with X-ray \cite{miller15} and $\gamma$-ray observations \cite{feldmann13}. }
\end{figure}

\subsubsection{For $r\lesssim 3$~kpc}

 In this region, the expressions of  $n_{{\rm{H_{2}}}}(r.z)$, $n_{{\rm{HI}}}(r.z)$, and $n_{{\rm{HII}}}(r.z)$ are considered same as the expressions for  Case \#2 in $r\lesssim 3$~kpc. It should be noted that in this case, similar to Case \#2, the  $n_{{\rm{H_{2}}}}(r.z)$ and $n_{{\rm{HI}}}(r.z)$ contain the combined contributions of CMZ and disk of GB (see the Eqs.~(14) and (17)) and $n_{{\rm{HII}}}(r.z)$ consists of the contribution of WIM (see Eq.~(18)).
 
\subsubsection{For $r > 3$~kpc} 
In this region of interest, we have constructed our desired density profiles with the radial ($n_{j} (r)$)  and vertical ($n_{j} (z) )$) distributions in the following way
  
 \begin{equation}
n_{j} (r, z) = N_{0_{j}} n_{j} (r) n_{j} (z)~\rm{cm^{-3}},
\end{equation} 
where, $j($ (= $\rm{H_{2}}$, HI, HII) denotes the particular component of hydrogen gas. Here, $N_{0_{j}}$ is the normalization constant that can be evaluated as, $N_{0_{j}} = 1/n_{j} (z = 0)$. 

Recent \textit{Fermi} data and hydrodynamical simulation indicate that a significant portion of diffuse gamma ray may come from the Milky Way Galaxy due to interaction of CR protons with the hot and ionized gas which can extend upto the virial radius of the Galaxy \cite{feldmann13}. Both of the \texttt{Adaptive Refinement Tree (ART)} code \cite{kravtsov1997,kravtsov02} and a high resolution cosmological, hydrodynamical simulation, containing the cosmological initial condition based on the \textit{Wilkinson Microwave Anisotropy Probe} \cite{spergel03} and a photochemical network, compute abundances of the various components of hydrogen gas \cite{feldmann13}. Depending on the simulation and the observations, the radial distribution of HII  and the z-dependence of ${\rm{H_{2}}}$, HI and HII have been shown in Ref.~\cite{feldmann13}. For the present work, we have considered the following z-dependence, based on the fitting of vertical distribution of Ref.~\cite{feldmann13}, of different components of the hydrogen gas

\begin{subequations}

\begin{equation}
n_{\rm{H_{2}}} (z) = 2.09 \times {\rm{exp}}\Bigg[   - \bigg( \frac{|z|}{0.29~{\rm{kpc}}}    \bigg)^{1.96} \Bigg]  ~{\rm{cm^{-3}}},
\label{eq:zgd1}
\end{equation}

\begin{equation}
n_{\rm{HI}}(z) = 2.09 \times {\rm{exp}}\Bigg[   - \bigg( \frac{|z|}{0.38~{\rm{kpc}}}    \bigg)^{1.76} \Bigg] ~ {\rm{cm^{-3}}},
\label{eq:zgd2}
\end{equation}
\rm{and}

\begin{eqnarray}
n_{\rm{HII}}(z) = 0.49 \times {\rm{exp}}\Bigg[  - \bigg( \frac{|z|}{0.40~{\rm{kpc}}}    \bigg)^{1.36} \Bigg]\nonumber \\
             + 7.05 \times 10^{-4} \times {\rm{exp}}\Bigg[  - \bigg( \frac{|z|}{9.17~{\rm{kpc}}}   \bigg) \Bigg] ~ {\rm{cm^{-3}}}.
 \label{eq:zgd3}             
\end{eqnarray}
\label{eq:zgd}
\end{subequations}

 We will use Eqs.~(\ref{eq:zgd}) in our chosen models to incorporate the z-dependence. We, now, considered different components of density profiles of hydrogen gas. In this case, we have , further, divided $r >3$~kpc region in two parts; one is $3~{\rm{kpc}} < r < 10.4~{\rm{kpc}}$ and  other one is $r \gtrsim 10.4~{\rm{kpc}}$.

 \subsubsubsection{For  $3~{\rm{kpc}} < r < 10.4~{\rm{kpc}}$}

  In this region, we have taken the radial distributions (i.e., $z = 0$) of Eqs.~(19), (20) and (22). We have used the vertical distributions, i.e. Eqs. (24), and the normalization procedure (discusses above) to construct the desired density profiles.

\textbf{\Large{$n_{{\rm{H_{2}}}}$}:}     The normalized density distribution can be written as
\begin{eqnarray}
n_{{\rm{H_{2}}}} (r, z)= && (0.5 \times 0.58~{\rm{cm^{-3}}}) \times \Bigg(  \frac{r}{8.5~{\rm{kpc}}}  \Bigg)^{-0.58} \nonumber \\
 && \times {\rm{exp}}\Bigg[  - \frac{(r - 4.5~{\rm{kpc}})^{2}  -  (4.0~{\rm{kpc}})^{2}   } {(2.9~{\rm{kpc}})^{2}}   \Bigg]   \nonumber \\
  &&  \times {\rm{exp}}\Bigg[   - \Big( \frac{|z|}{0.29}    \Big)^{1.96} \Bigg].
 \end{eqnarray}

\textbf{\Large{$n_{{\rm{HI}}}$}:}  The normalized density distribution of HI is constructed as 
\begin{eqnarray}
n_{{\rm{HI}}}(r,z) = &&\frac{(0.340~{\rm{cm^{-3}}} )}{(\alpha_{h} (r))^{2}}  \times {\rm{exp}}\Bigg[   - \Big( \frac{|z|}{0.38~{\rm{kpc}}}    \Big)^{1.76} \Bigg)      \Bigg] \nonumber \\
&&+ \frac{(0.226~{\rm{cm^{-3}}} )}{(\alpha_{h} (r))} \times {\rm{exp}}\Bigg[  - \Big( \frac{|z|}{0.38~{\rm{kpc}}}    \Big)^{1.76} \Bigg] \nonumber \\
&&  \times \Bigg\{\Bigg[  1.745 - \frac{1.289}{\alpha_{h} (r))} \Bigg] +  \Bigg[ 0.473 - \frac{0.070}{\alpha_{h} (r))} \Bigg] \nonumber \\
 && +  \Bigg[  0.283 - \frac{0.142}{\alpha_{h} (r))} \Bigg] \Bigg\}.
\end{eqnarray}
 where,
\begin{equation}
\begin{rcases}
 \alpha_{h} (r) &= 1.0,~ {\rm{For,~ r \leq 8.5~kpc}}  \\
                                      &= \frac{r}{8.5 ~{\rm{kpc}}}, ~ {\rm{For,~ r > 8.5~kpc}}.\\
\end{rcases}
\end{equation}
 \\

\textbf{\Large{$n_{{\rm{HII}}}$}:}
In this case, the normalized density profile is considered as, 
\begin{widetext}
 \begin{eqnarray}
 n_{{\rm{HII}}}(r,z) =&& \Bigg( (0.0237~{\rm{cm^{-3}}} )~{\rm{exp}}\Bigg[-\frac{r^{2} - (8.5~{\rm{kpc}})^{2} }{(37.0~{\rm{kpc}})^{2}}\Bigg] \nonumber \\
&&+ (0.0013~{\rm{cm^{-3}}} )  \times  {\rm{exp}}\Bigg[-\frac{(r- 4.0~{\rm{kpc}})^{2} - (4.5~{\rm{kpc}})^{2}    }{(2.0~{\rm{kpc}})^{2}} \Bigg] \Bigg) \nonumber \\
&& \times \frac{1.0}{0.491} \Bigg( 0.49 \times {\rm{exp}}\Bigg[  - \bigg( \frac{|z|}{0.40~{\rm{kpc}}}    \bigg)^{1.36} \Bigg] + 7.05 \times 10^{-4} \times {\rm{exp}}\Bigg[  - \bigg( \frac{|z|}{9.17~{\rm{kpc}}}   \bigg) \Bigg]  \Bigg).
\end{eqnarray}
 \end{widetext}

 \subsubsubsection{For  $ r \gtrsim  10.4~{\rm{kpc}}$}

 In this region, the density distributions of molecular, atomic and ionized components of hydrogen gas are obtained by using the following relation
 
 \begin{equation}
n_{{\rm{H}}}(r) - n_{{\rm{HII}}} (r)  = 2n_{{\rm{H_{2}}}}(r) + n_{{\rm{HI}}}(r).
\label{eq:211}
\end{equation}
Ref.~\cite{miller15} provides the density of  hydrogen gas (i.e., $n_{\rm{H}}$) which is obtained from the X-ray spectroscopy by measuring the OVII and OVIII absorption lines at zero redshift or following the emission line in the blank sky spectrum. We also assumed that $n_{{\rm{H_{2}}}}(r) = n_{{\rm{HI}}}(r) =  \Big(n_{{\rm{H}}}(r) - n_{{\rm{HII}}}(r)\Big )/3$ and $n_{{\rm{HII}}} (r)$ is obtained from Ref. \cite{feldmann13} (see below and the above discussion started for the section of $r >3$~kpc ).

\textbf{\Large{$n_{{\rm{H_{2}}}}$}:}     The normalized density distribution can be written as

\begin{eqnarray}
n_{{\rm{H_{2}}}}(r.z) = &&\frac{\Big(n_{{\rm{H}}}(r) - n_{{\rm{HII}}} (r)\Big )}{3} \nonumber \\
&& \times {\rm{exp}}\Bigg[  - \Big( \frac{|z|}{0.29~{\rm{kpc}}}    \Big)^{1.96} \Bigg] ~{\rm{cm^{-3}}}.
\label{eq:gd21}
\end{eqnarray}

\textbf{\Large{$n_{{\rm{HI}}}$}:}  The normalized density distribution of HI is constructed as 

\begin{eqnarray}
n_{{\rm{HI}}}(r.z) = && \frac{\Big(n_{{\rm{H}}}(r) - n_{{\rm{HII}}} (r)\Big )}{3} \nonumber \\
&&\times {\rm{exp}}\Bigg[   - \Big( \frac{|z|}{0.38 ~{\rm{kpc}}}    \Big)^{1.76} \Bigg]  ~{\rm{cm^{-3}}},
\label{eq:gd2&&2}
\end{eqnarray} 
\\
\textbf{\Large{$n_{{\rm{HII}}}$}:}
In this case, the normalized density profile is considered as,

 \begin{equation}
n_{{\rm{HII}}}(r.z) = n_{{\rm{HII}}} (r) \times \Big(n_{{\rm{HII}}}(z)/ n_{{\rm{HII}}}(z = 0)\Big),
\label{eq:gd221}   
\end{equation}   
where, data for $n_{{\rm{HII}}}(r)$ is obtained from Ref.~\cite{feldmann13}
(see the discussion above).

In Figs.~3(a,b,c), we have shown the density profiles of molecular, atomic and ionized hydrogen gas. In the $r \lesssim 3$~kpc region, we have considered a co-ordinate transformation, i.e. from Cartesian ($x,y,z$) to cylindrical ($r, \theta, z$) and taken the azimuthal average of Eqs.~(14), (17) and (18) to obtain $n_{{\rm{H_{2}}}}(r.z)$, $n_{{\rm{HI}}}(r.z)$, and $n_{{\rm{HII}}}(r.z)$. In Figs.~3(a,b), a gap in the range 1-3 kpc has been observed due to effect of Galactic bar (as explained in Case \#2).  Here, the vertical distributions, in the range above 3 kpc, are obtained from Ref.~\cite{feldmann13}. The radial distribution, i.e., $r\gtrsim10.4$~kpc is obtained from Refs.~\cite{miller15, feldmann13}. To plot the density profiles, proper interpolation, whenever needed, is done between the data of different regions.

\section{\label{sec:results}Results}

In this section, we have studied the variation of $z_{t}$ for each density profile assuming $R_{\rm{max}}$ to be 20 kpc. In the present analysis, we have considered plain diffusion model with the assumption $L = 3z_{t}$. In each of the case, we will fit the observed CR data in the following procedure

a)  First we consider the relation between $z_{t}$ and  $B_0^{\rm{turbulent}}$ which is obtained under the condition that the CR electron flux  from different CR electron proapagation models reproduces the observed synchrotron spectrum at 408 MHz \cite{bernardo13}. Then,  we will fit $^{10}$Be/$^{9}$Be to get an estimate of $z_{t}$ and the corresponding $B_0^{\rm{turbulent}}$ is obtained from Ref.~\cite{bernardo13}.  Simultaneously, proton (p) and helium (He) are also fitted to obtain the other parameters.\\

b) We will then change $D_{0}$ to fit B/C by keeping fixed $z_{t}$, $B_0^{\rm{turbulent}}$ and other parameters as obtained in (a). If needed, the parameters (except $z_{t}$ and $B_0^{\rm{turbulent}}$) are fine tuned to improve the fitting of B/C. Simultaneously , we also check the fitting of $^{10}$Be/$^{9}$Be, p, He and antiproton ($\bar{p}$) for consistency.

\subsection{Case~1} 

\begin{figure*}[!ht]
\centering
\mbox{
\includegraphics[width=0.45\textwidth,clip,angle=0]{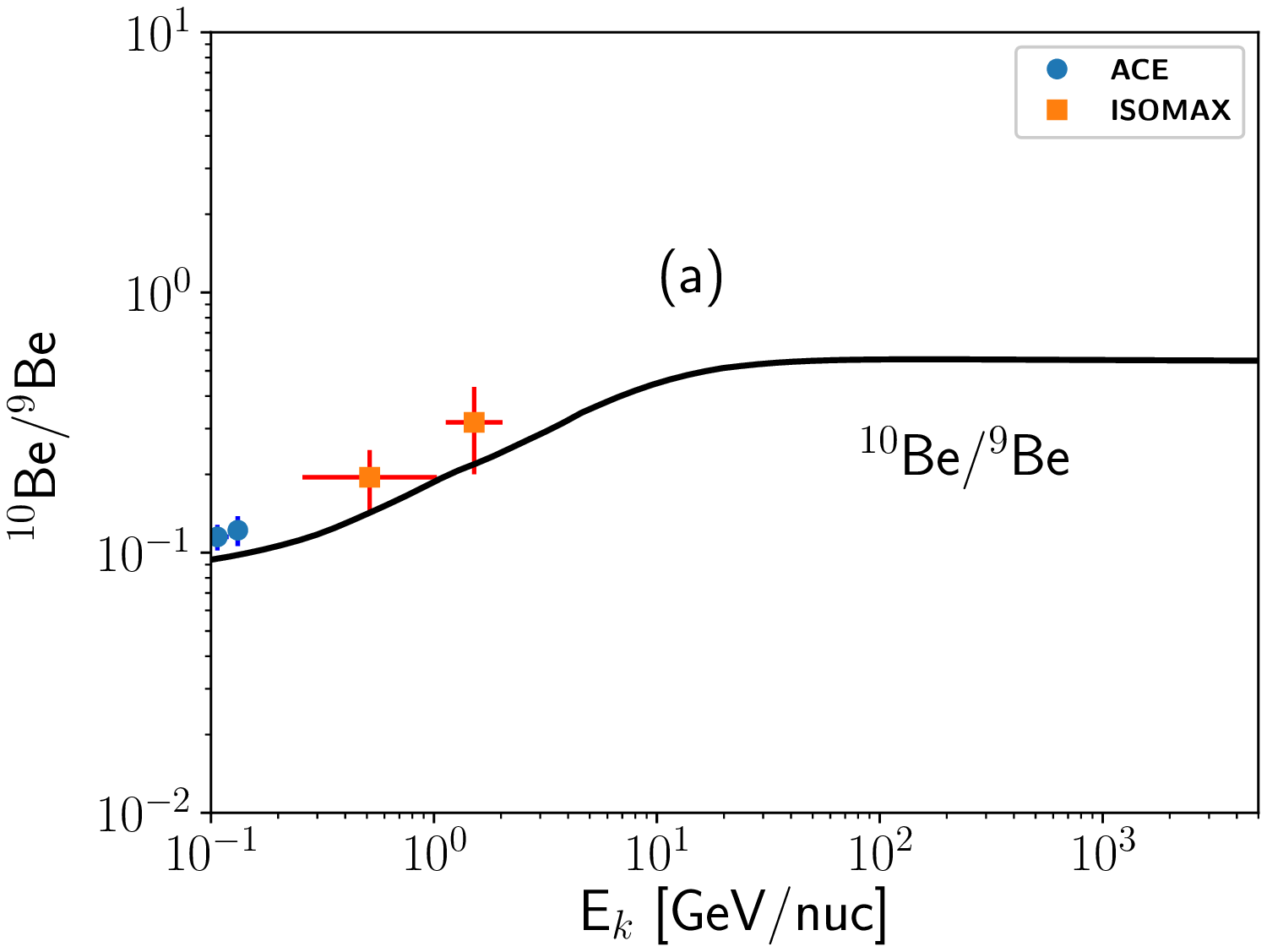}

\includegraphics[width=0.45\textwidth,clip,angle=0]{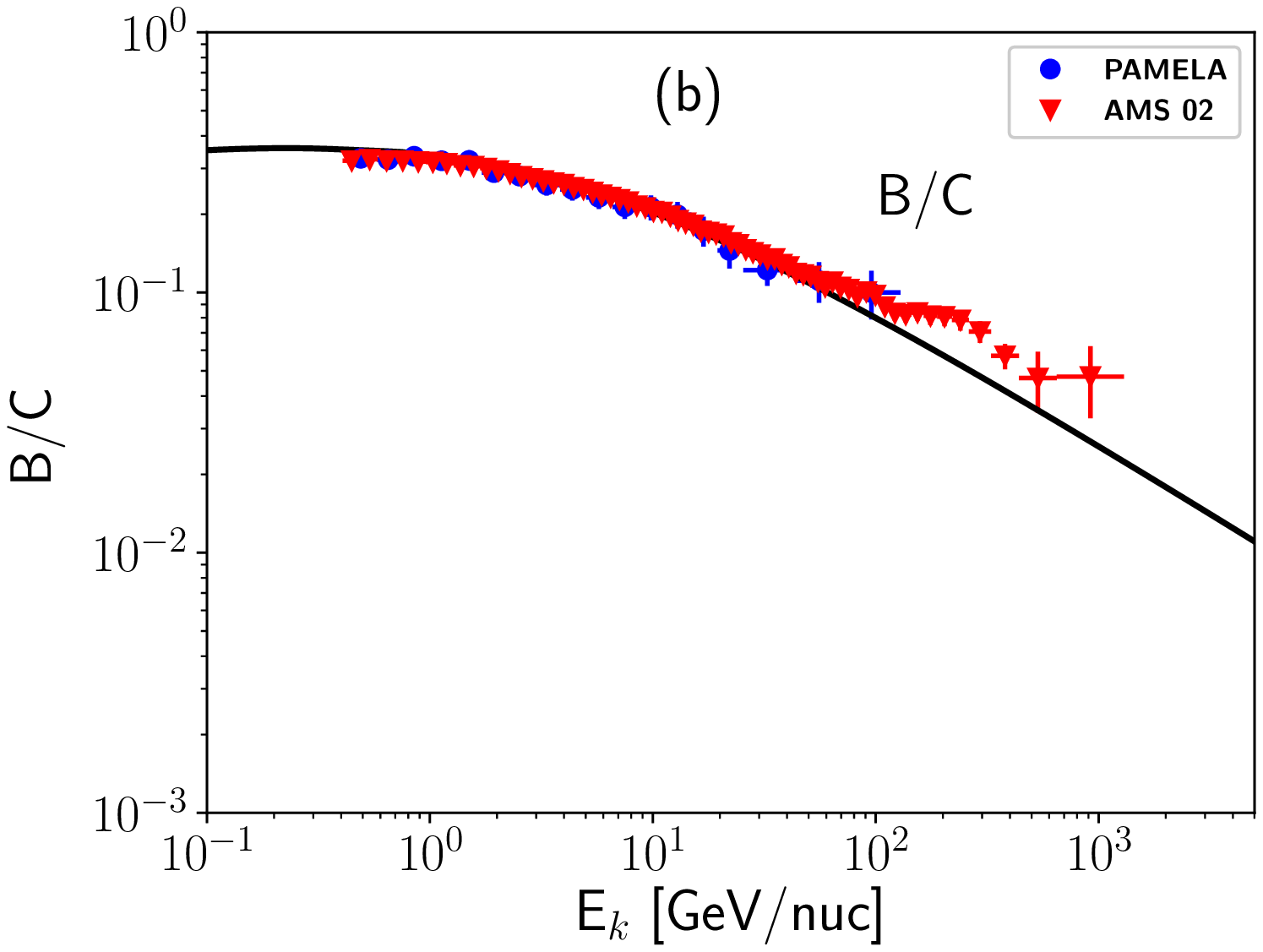}
}

\mbox{
\includegraphics[width=0.45\textwidth,clip,angle=0]{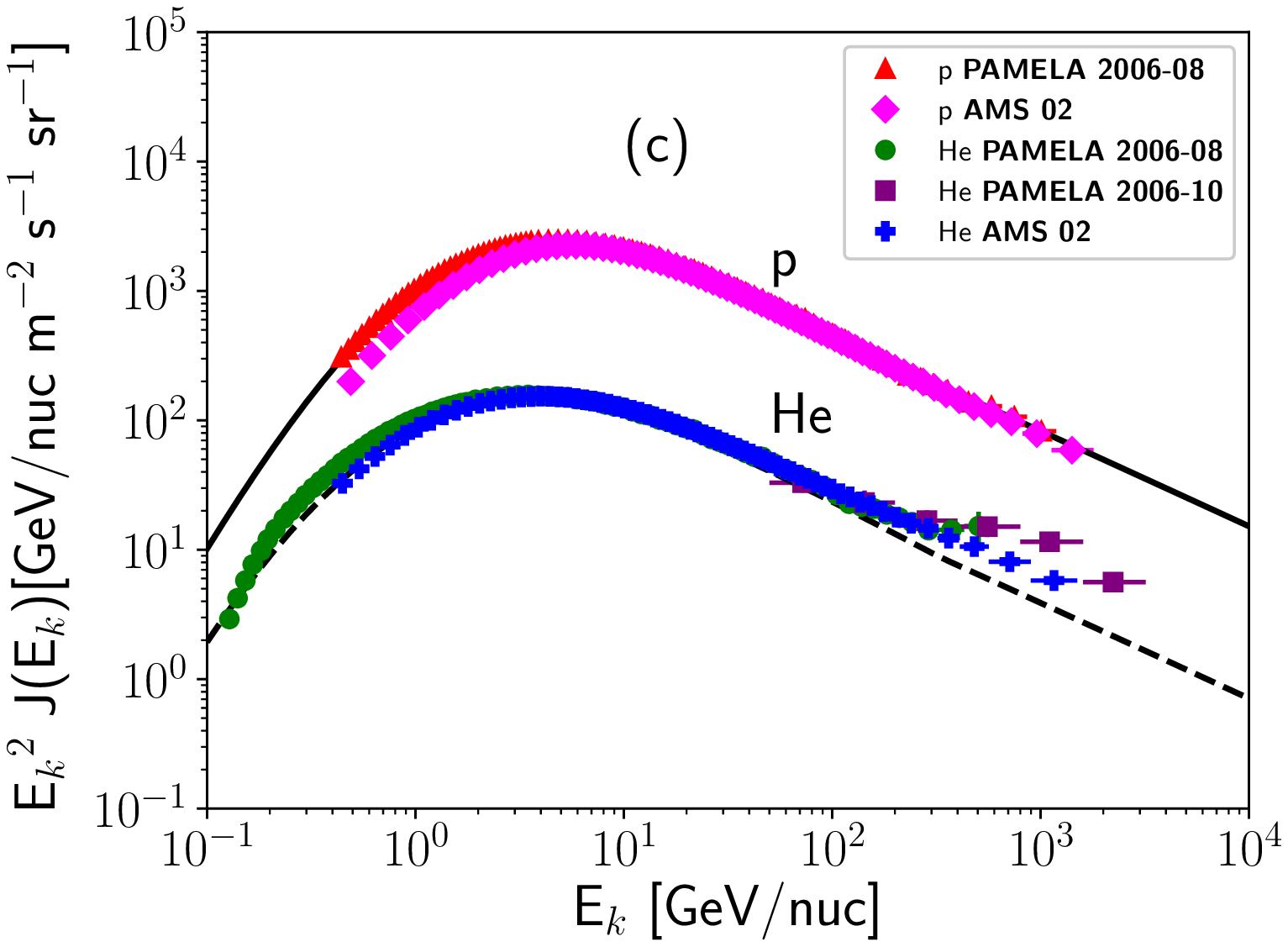}

\includegraphics[width=0.45\textwidth,clip,angle=0]{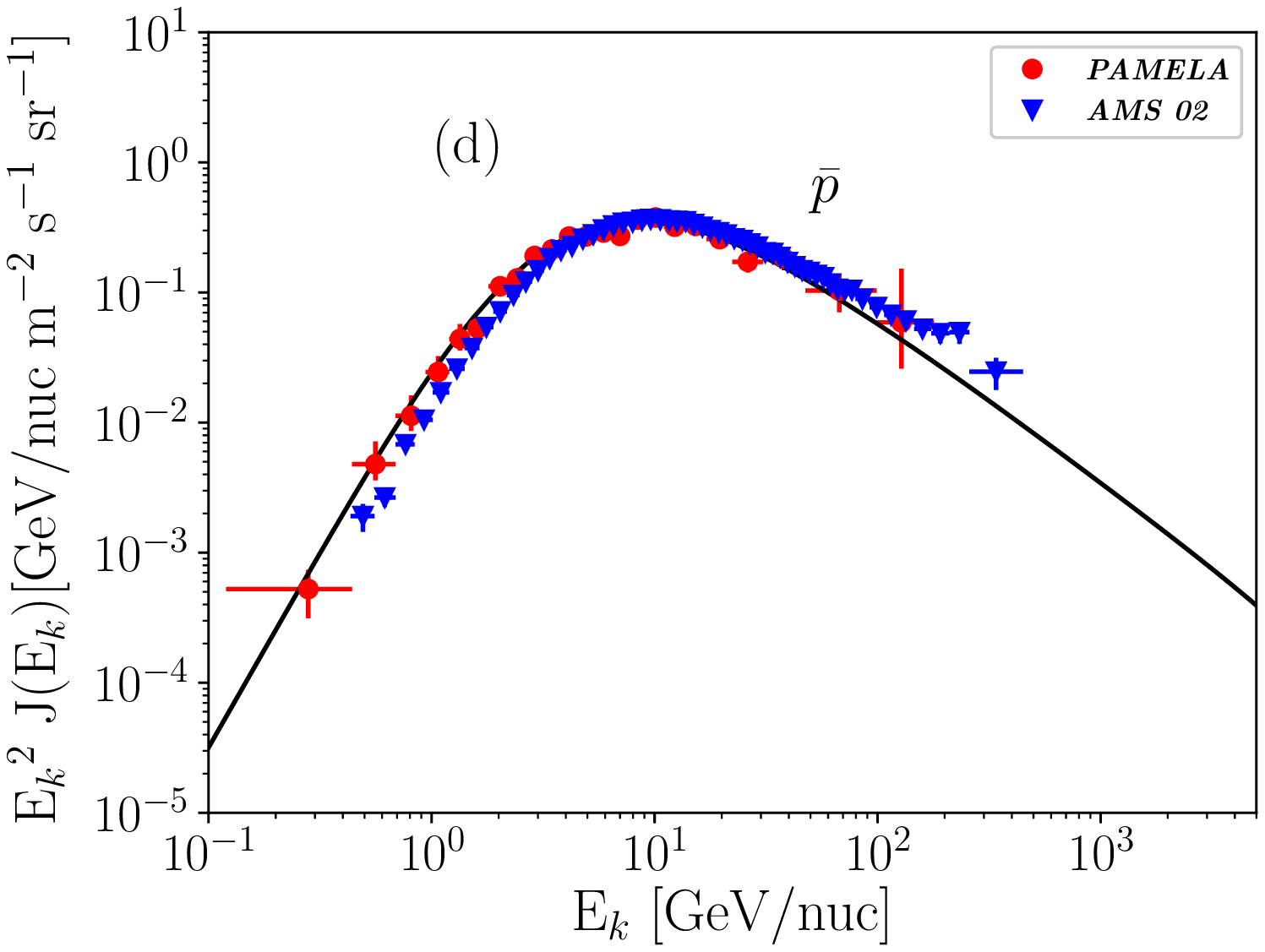}

}

\caption{\label{fig:fig2} (Color online)  In the figure (a) $^{10}$Be/$^{9}$Be ratio, obtained by using Case~\#1 in the \texttt{DRAGON} code, is plotted with the ACE \cite{ace} and ISOMAX \cite{isomax} data, (b) B/C ratio is plotted with the PAMELA \cite{pamelabc} and AMS 02 \cite{amsbc}, (c)  p and He fluxes are plotted with the PAMELA \cite{pamelap,pamelahe2} and AMS 02 \cite{amsp,amshe} and (d) $\bar{p}$ flux with PAMELA \cite{pamelapbar} and AMS 02 \cite{amspbar} data are shown. Here, $\phi=0.58$~GV. The parameter set to fit CR spectra has been tabulated in Table II.}

\end{figure*}

\begin{table}[!ht]
\caption{\label{tab:tab1} Various models and parameter values selected in \texttt{DRAGON} code to fit CR spectra shown in Figs.~\ref{fig:fig2}(a,b,c,d) are listed here.  The density profiles of Case~\#1 have been selected for the study. }
\begin{ruledtabular}
\begin{tabular}{ |p {5 cm}| l | }

Model/Parameter  &   Option/Value \\
 \hline
  $R_{\rm{max}}$ 			                                                                      &     20.0 ~ kpc \\  
  $L$                      																				&      6.0~ kpc  \\
  Gas density type																					&        Case \verb+#1+ \\
  
  Source Distribution          															 		&   Ferriere \\
  Diffusion type																						&        Exp  (see Eq.~(1))       \\
  $D_0$       																					&    $2.0 \times 10^{28}$~ $\rm{cm^2/s}$   \\
  $\rho_0$																						&     3.0~GV            \\
  $\delta$																		                &       0.54          	\\
  $z_t$ 																							&		2.0~kpc			\\
  $\eta$   																						&		-0.40		    \\
  $v_A$   																					&			0.0		        \\
  Magnetic field type																&	    Pshirkov			      \\
  $B_0^{\rm{disc}}$ 															&      $2.0 \times 10^{-6}$ ~ Gauss      \\
  $B_0^{\rm{halo}}$            												&    $4.0 \times 10^{-6}$ ~ Gauss                \\
  $B_0^{\rm{turbulent}}$   											 &       $9.65 \times 10^{-6}$ ~ Gauss             \\
  First injection slope ($\alpha_0$) 									&  2.32          \\
  Position of first break (rigidity)	                  					 &    330 ~GV      \\
  Second injection slope ($\alpha_1$)         						  &      2.20     \\
    
\end{tabular}
\end{ruledtabular}
\end{table}

\noindent Here, we consider the density profiles of  Case \#1.  With the input selections in the \texttt{DRAGON} code, as discussed in Sec.III, we fit the observed data set of $^{10}$Be/$^{9}$Be and boron to carbon (B/C) flux ratio by following the procedure mentioned above. Furthermore, we have also fitted the observed fluxes ($J(E_{k})$) of p, He and $\bar{p}$. We have chosen, $ D_{0}= 2.0 \times 10^{28} {\rm{cm^{2}}/s}$, $\delta = 0.54$, $\eta = -0.40$, $z_{t} = 2~{\rm{kpc}}$, $B_0^{\rm{turbulent}} = 9.65 \times 10^{-6}$~Gauss, $B_0^{\rm{disc}} = 2 \times 10^{-6}$~Gauss, $B_0^{\rm{halo}} = 4 \times 10^{-6}$~Gauss and $\phi = 0.58$~GV to tune the data set of $^{10}$Be/$^{9}$Be, B/C, p, He and $\bar{p}$. The values of the disc component and the halo component of GMF are obtained on the basis of Faraday rotation measurements \cite{bernardo13, pshirkov11}. Our $\delta$ value is higher than that reported in Ref.~\cite{amsbc}. We find $\delta = 0.54$ gives a reasonably good fit to the CR data in the energy range of 0.1 to 1000 GeV/nuc. The models and parameter values used in this case are listed in Table~\ref{tab:tab1}.

 In Fig.\ref{fig:fig2}(a), we have plotted the $^{10}$Be/$^{9}$Be flux ratios with the observed data set\footnote{All the data are taken from the cosmic ray database \cite{crd}.} of $^{10}$Be/$^{9}$Be from ACE \cite{ace} and ISOMAX \cite{isomax}. Similarly, the B/C flux ratios with the observed data set for PAMELA \cite{pamelabc} (the data set is for 2006-2008 unless otherwise specified) and AMS 02 \cite{amsbc} are also plotted in Fig.\ref{fig:fig2}(b). Proton (p), helium (He) and antiproton ($\bar{p}$)  fluxes are also plotted with PAMELA and AMS 02 data in Fig.\ref{fig:fig2}(c) and Fig.\ref{fig:fig2}(d), respectively. We would like to add a note that the fitting of $^{10}$Be/$^{9}$Be would be improved if we would decrease $z_{t}$ below 2 kpc. But $z_{t} < 2$~kpc is strongly disfavored (with $3 \sigma$ confidence level) by the measurements of radio maps of selected regions \cite{bernardo13,bringmann12}. Our chosen parameter set fits the observed CR data quite well in the whole energy range.

\subsection{Case~2}

\begin{figure*}[!h]

\centering
\mbox{

\includegraphics[width=0.45\textwidth,clip,angle=0]{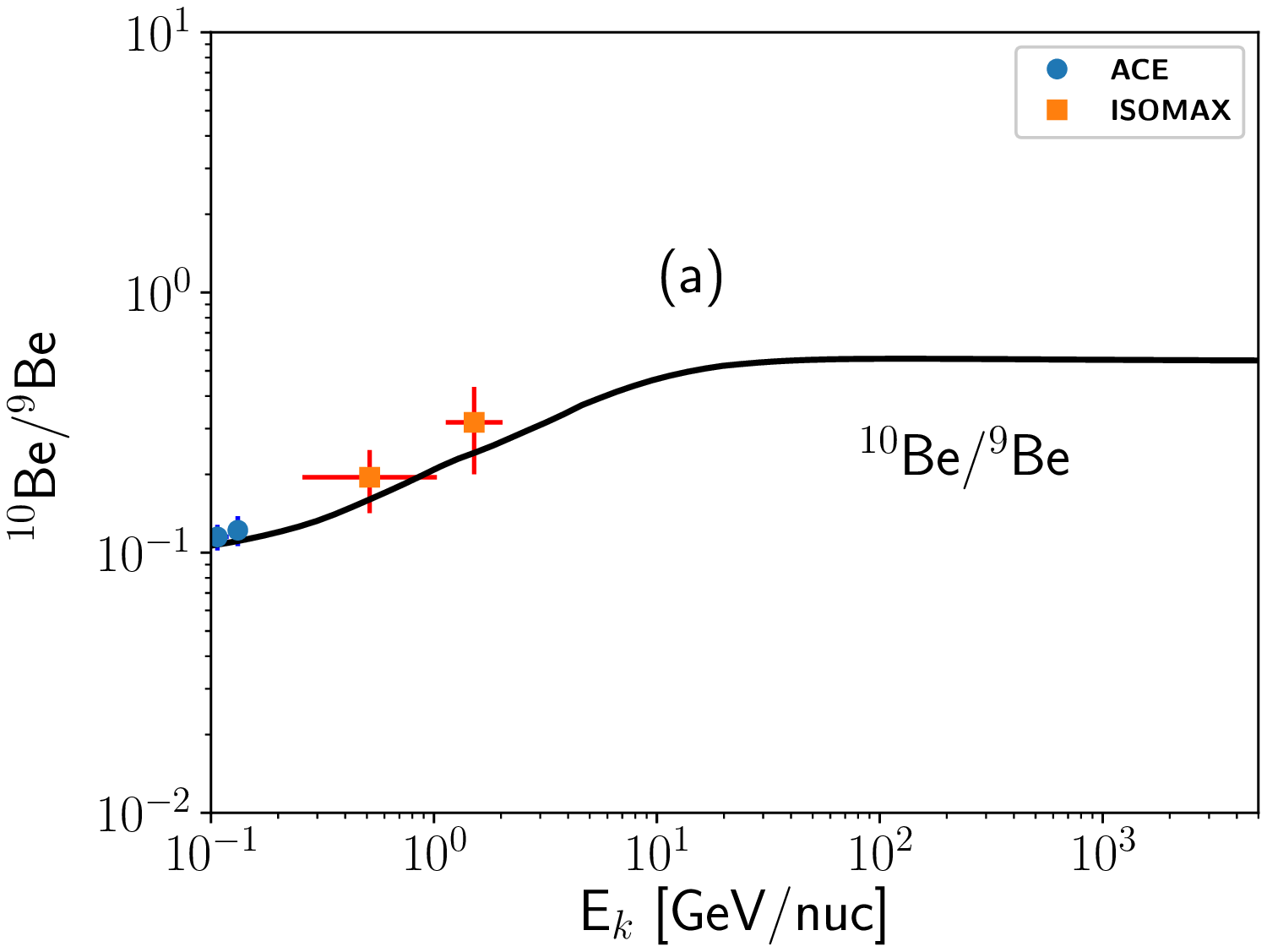}

\includegraphics[width=0.45\textwidth,clip,angle=0]{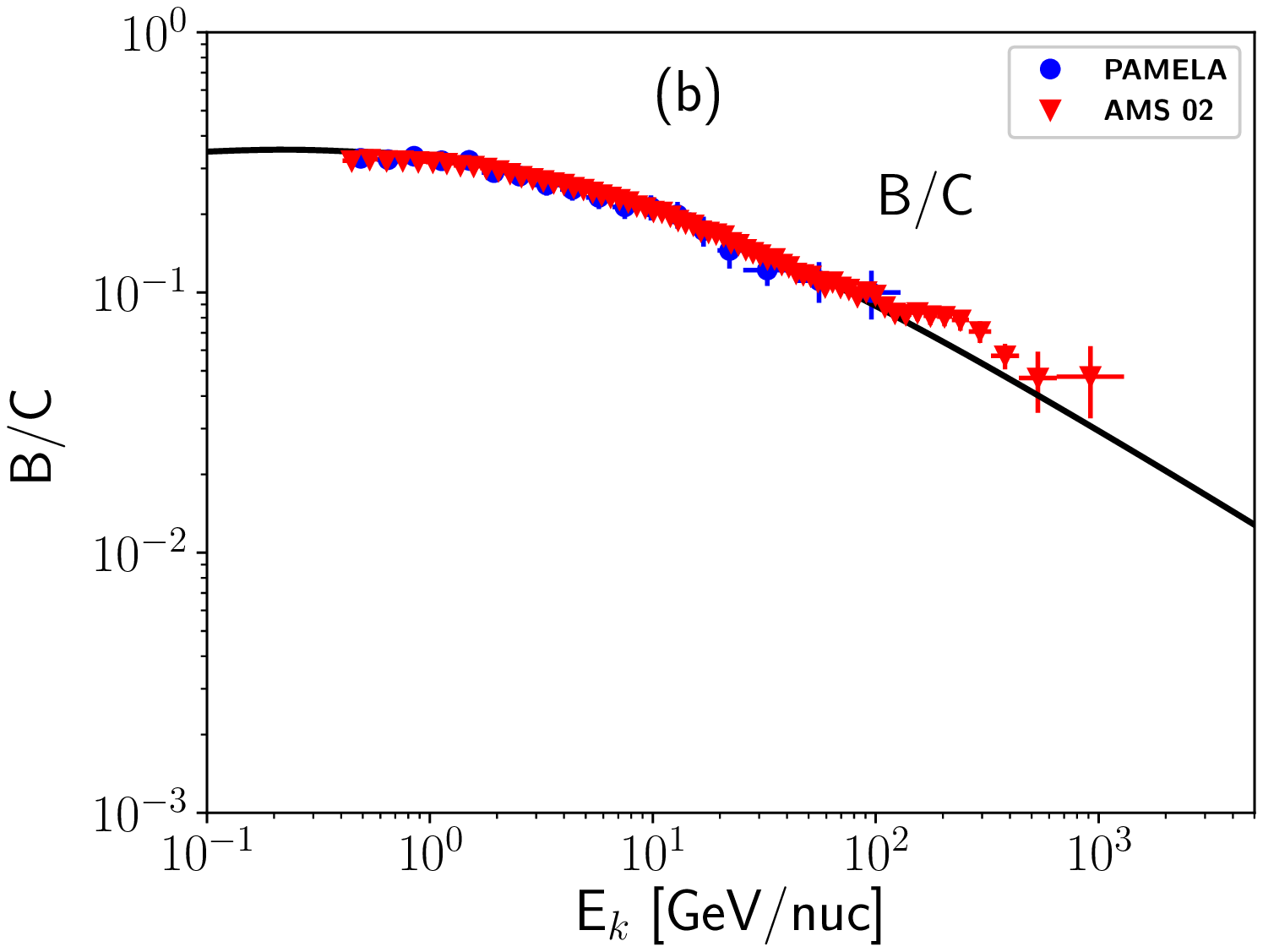}
}

\mbox{
\includegraphics[width=0.45\textwidth,clip,angle=0]{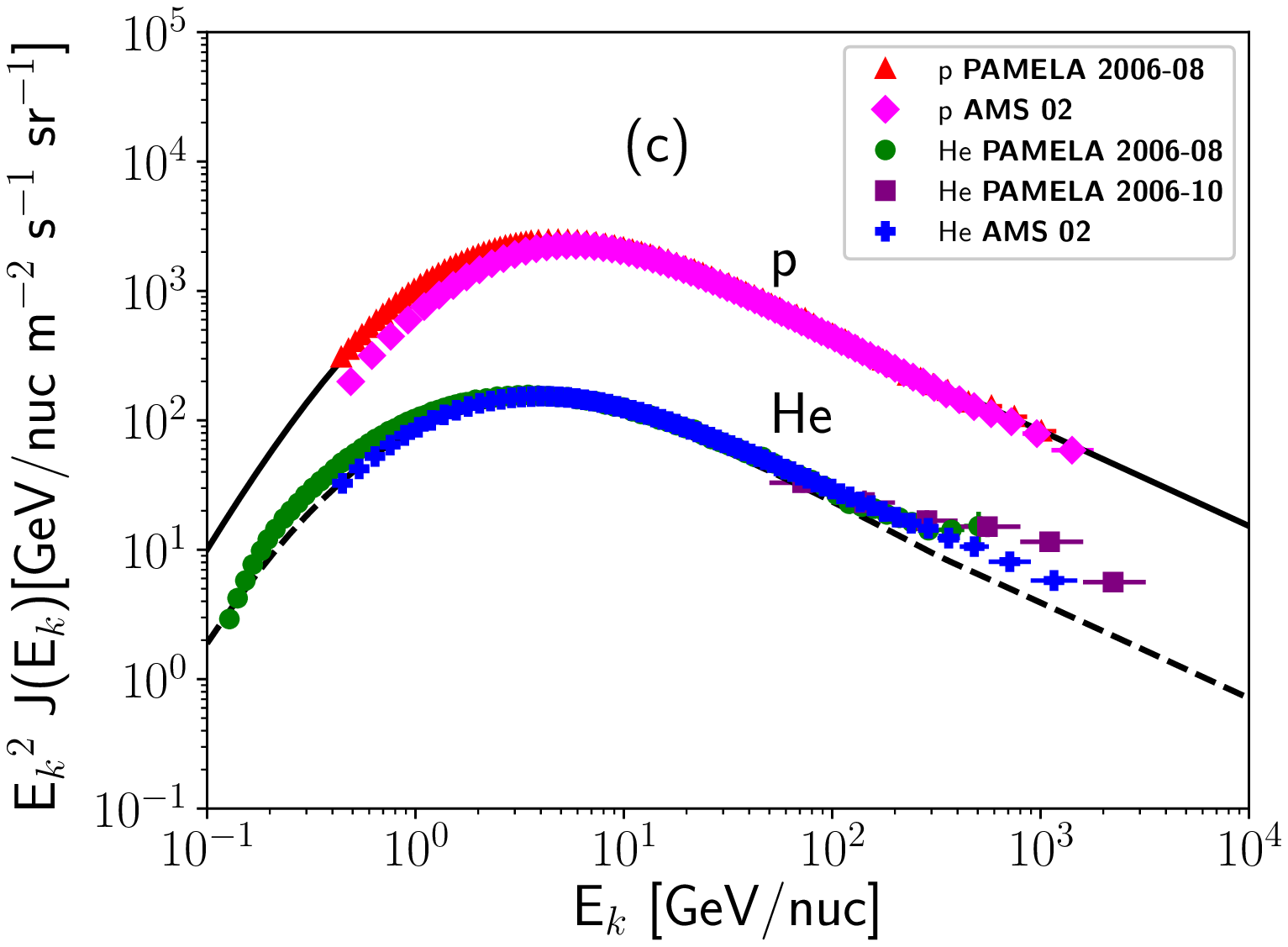}

\includegraphics[width=0.45\textwidth,clip,angle=0]{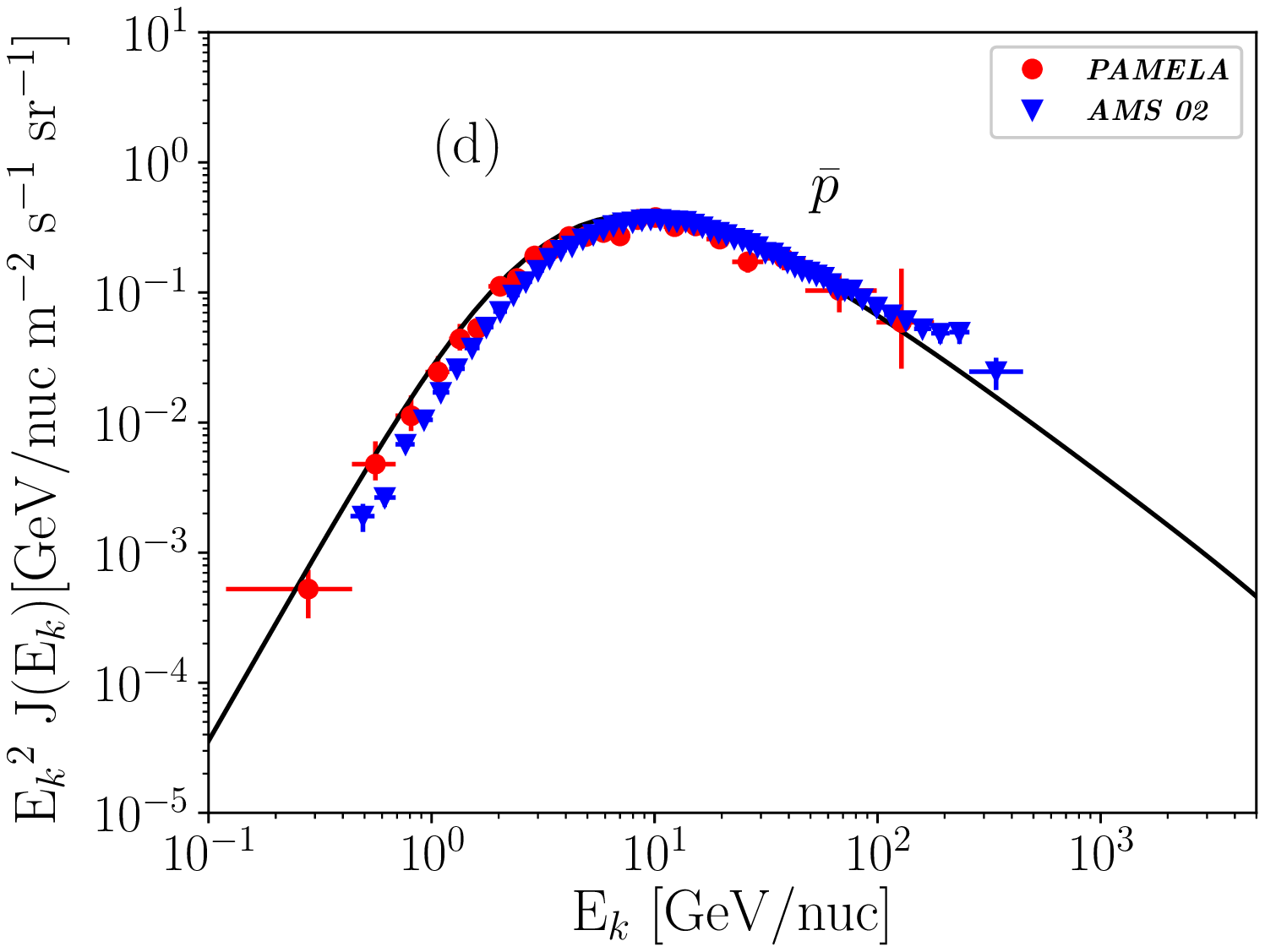}
}
\caption{\label{fig:fig5} (Color online)  In the figure (a) $^{10}$Be/$^{9}$Be ratio, obtained by using Case~\#2 in the \texttt{DRAGON} code, is plotted with the ACE \cite{ace} and ISOMAX \cite{isomax} data, (b) B/C ratio is plotted with the PAMELA \cite{pamelabc} and AMS 02 \cite{amsbc}, (c)  p and He fluxes are plotted with the PAMELA \cite{pamelap,pamelahe2} and AMS 02 \cite{amsp,amshe} and (d) $\bar{p}$ flux with PAMELA \cite{pamelapbar} and AMS 02 \cite{amspbar} data are shown. Here, $\phi=0.58$~GV. The parameter set to fit CR spectra has been tabulated in Table III. The significant parameter values for these fitted spectra are $z_{t} = 2.2$~kpc, $L = 6.6$~kpc, $D_0 =2.9 \times 10^{28}$~ $\rm{cm^2/s}$ and $B_0^{\rm{turbulent}} =9.30 \times 10^{-6}$ ~ Gauss.}

\end{figure*}

\begin{table}[!ht]
\caption{\label{tab:tab2} Various models and parameter values selected in \texttt{DRAGON} code to fit CR spectra shown in Figs.~5(a,b,c,d) are listed here. The density profiles of Case~\#2 have been selected for the study. Significant changes (in comparison to Table~\ref{tab:tab1}) are shown in bold text. }
\begin{ruledtabular}
\begin{tabular}{ |p {5 cm}| l | }

Model/Parameter  &   Option/Value \\
 \hline
  $R_{\rm{max}}$																		 			   &     20.0 ~ kpc \\  
   $L$                      																				&      \textbf{6.6~ kpc}  \\
  Gas density type																		&      \bf{ Case \verb+#2+  }\\
  
  Source Distribution          															 &   Ferriere \\
  Diffusion type																				&        Exp  (see Eq.~(1))         \\
  $D_0$       																	&    $\bf{2.9 \times 10^{28}}$~ $\bf{\rm{cm^2/s}}$  \\
  $\rho_0$																						&     3.0~GV            \\
  $\delta$																		                &       0.54          	\\
  $z_t$ 																							&		\textbf{2.2~kpc}			\\
  $\eta$   																						&		-0.40		    \\
  $v_A$   																					&			0.0		        \\
  Magnetic field type																&	    Pshirkov			      \\
  $B_0^{\rm{disc}}$ 															&      $2.0 \times 10^{-6}$ ~ Gauss      \\
  $B_0^{\rm{halo}}$            												&    $4.0 \times 10^{-6}$ ~ Gauss                \\
  $B_0^{\rm{turbulent}}$   											 &      $\bf{9.30 \times 10^{-6}}$ ~\bf{ Gauss}             \\
  First injection slope ($\alpha_0$) 									&  2.32          \\
  Position of first break (rigidity)	                  					 &    330 ~GV      \\
  Second injection slope ($\alpha_1$)         						  &      2.20     \\
    
\end{tabular}
\end{ruledtabular}
\end{table}

In this case, we have considered the density profiles of Case~\#2. We follow the same procedure to fit the observed CR data as used in Case~1. In Table~\ref{tab:tab2}, we have listed various models and parameter values used in the \texttt{DRAGON} code to fit the CR spectra along with the chosen density profile. We need $z_{t} = 2.2$~kpc, $L = 6.6$~kpc, $D_0 =2.9 \times 10^{28}$~ $\rm{cm^2/s}$ and $B_0^{\rm{turbulent}} =9.30 \times 10^{-6}$ ~ Gauss to fit $^{10}$Be/$^{9}$Be, B/C, p, He, and $\bar{p}$ spectra for Case~\#2. The other parameters remain same as those mentioned in Table~\ref{tab:tab1}.

\subsection{Case~3}

\begin{figure*}[!h]
\centering
\mbox{
\includegraphics[width=0.45\textwidth,clip,angle=0]{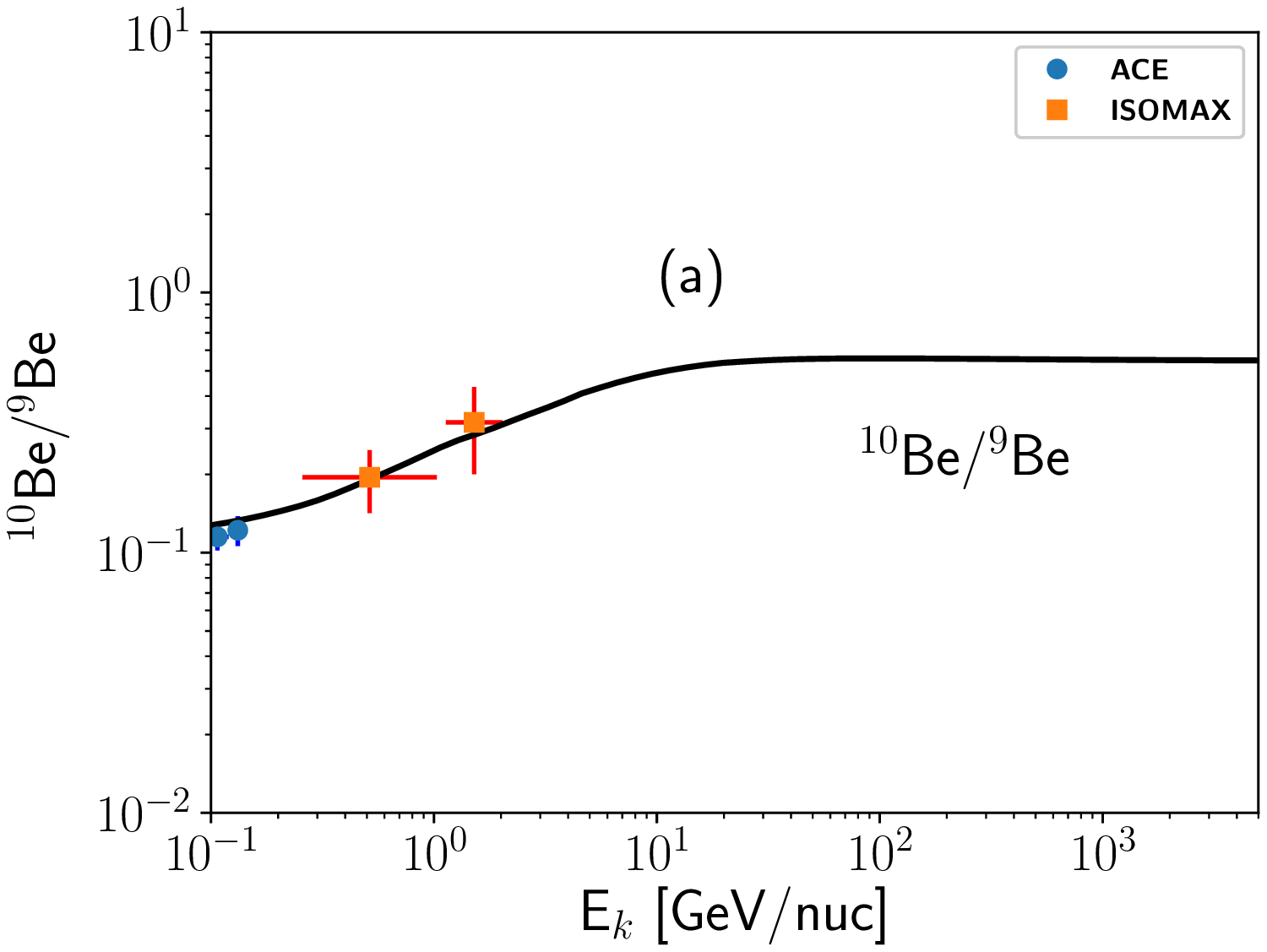}
\includegraphics[width=0.45\textwidth,clip,angle=0]{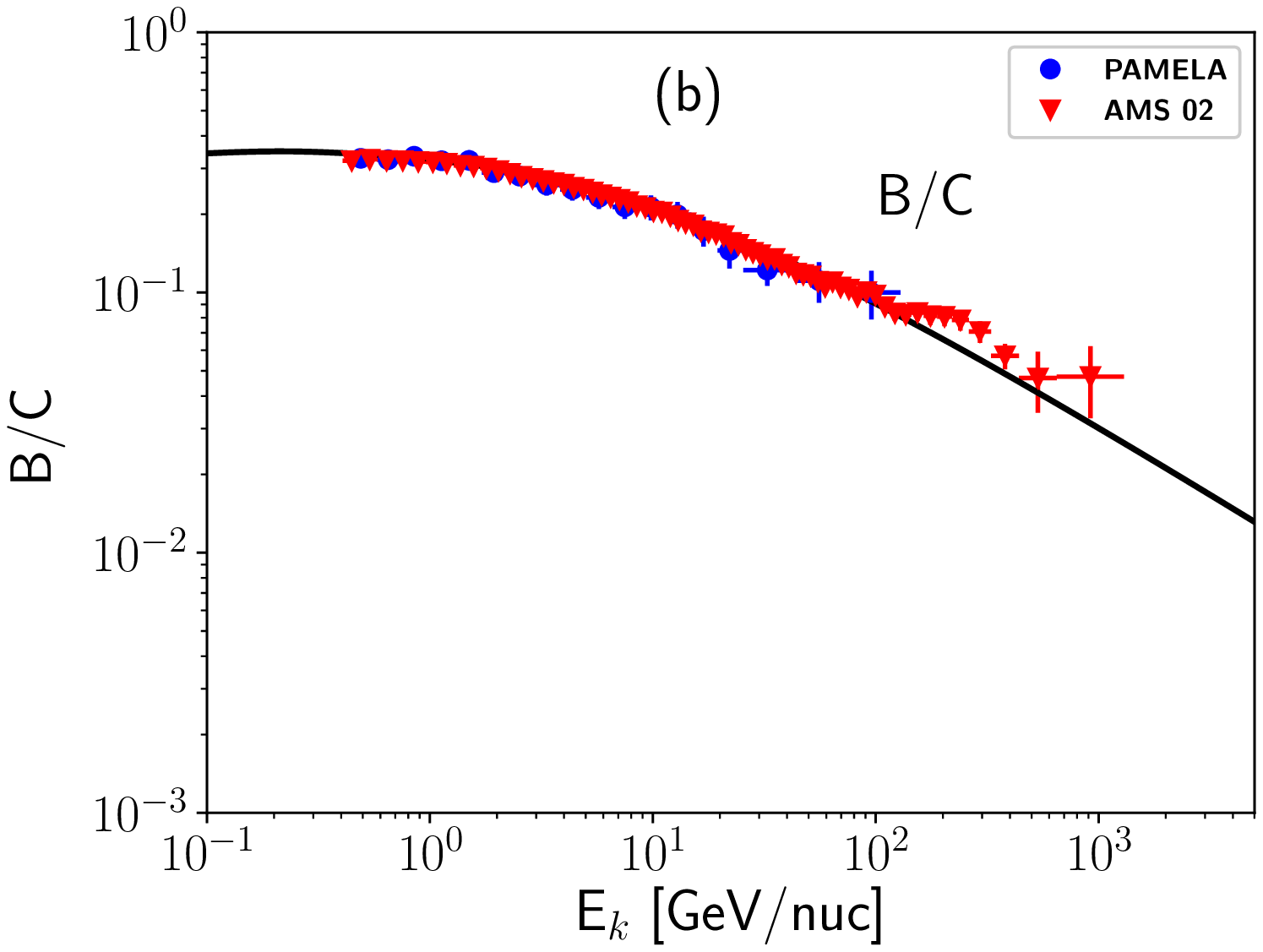}
}

\mbox{
\includegraphics[width=0.45\textwidth,clip,angle=0]{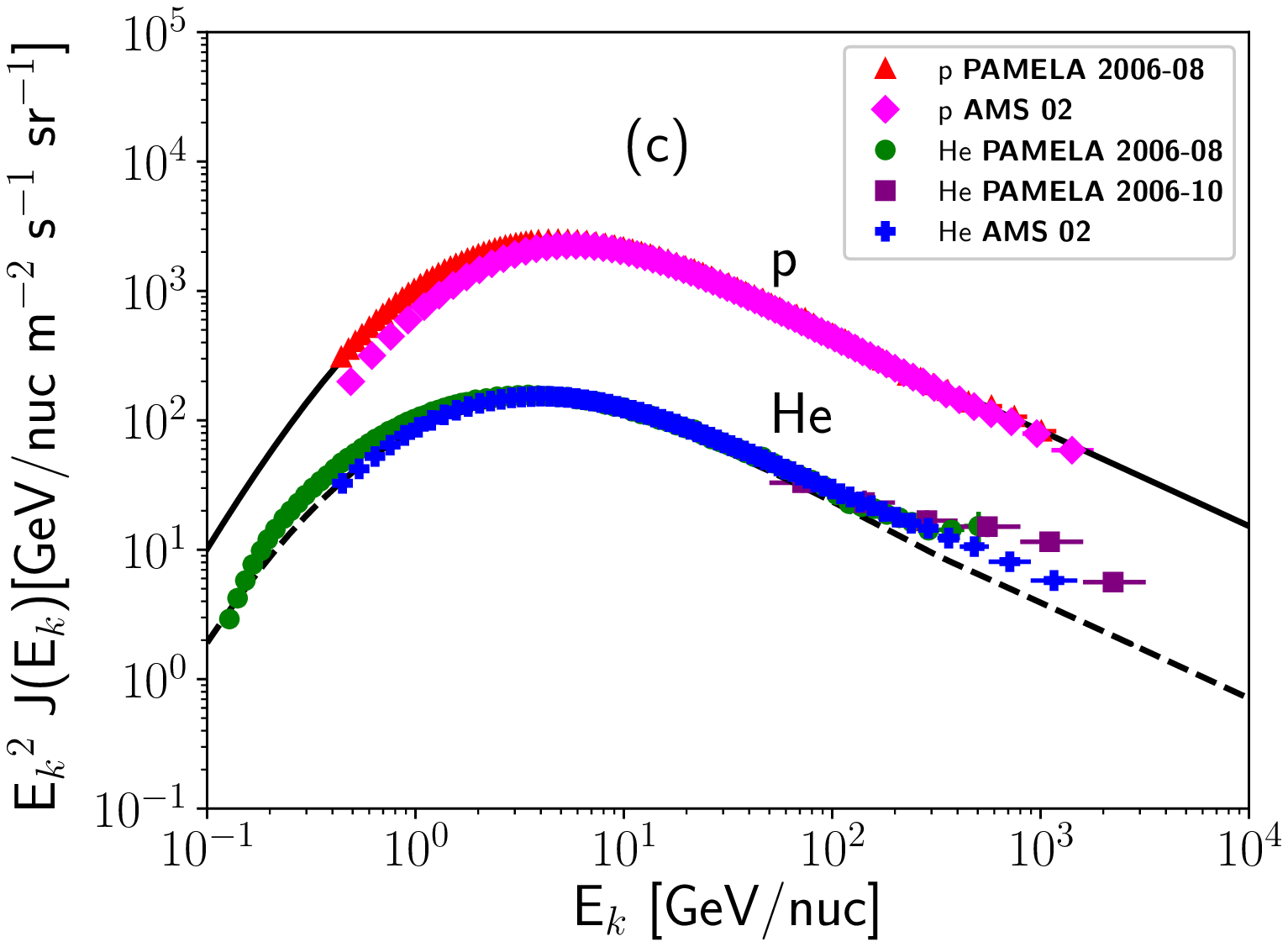}

\includegraphics[width=0.45\textwidth,clip,angle=0]{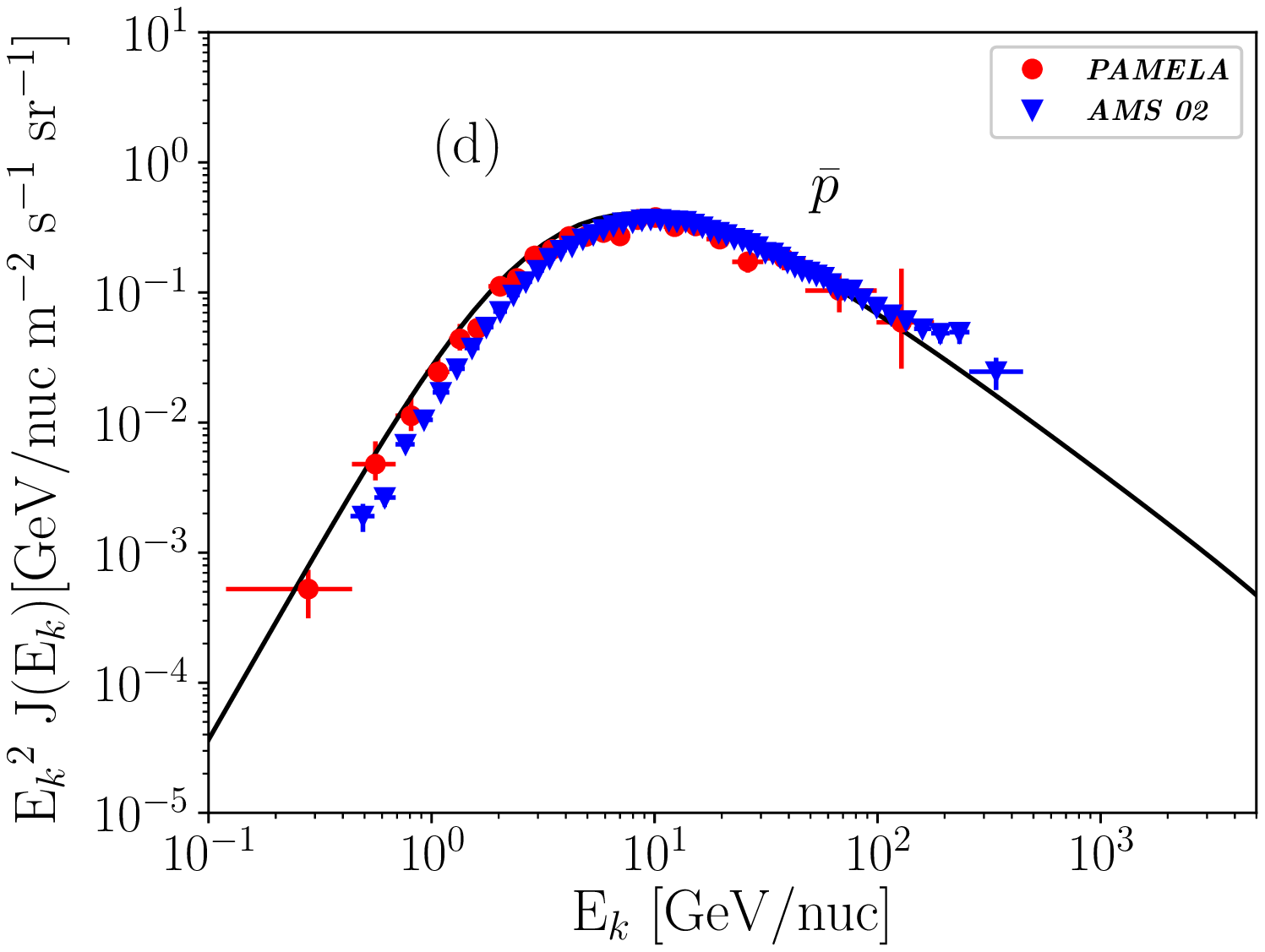}
}

\caption{\label{fig:fig6} (Color online)  In the figure (a) $^{10}$Be/$^{9}$Be ratio, obtained by using Case~\#3 in the \texttt{DRAGON} code, is plotted with the ACE \cite{ace} and ISOMAX \cite{isomax} data, (b) B/C ratio is plotted with the PAMELA \cite{pamelabc} and AMS 02 \cite{amsbc}, (c)  p and He fluxes are plotted with the PAMELA \cite{pamelap,pamelahe2} and AMS 02 \cite{amsp,amshe} and (d) $\bar{p}$ flux with PAMELA \cite{pamelapbar} and AMS 02 \cite{amspbar} data are shown. Here, $\phi=0.58$~GV. The parameter set to fit CR spectra has been tabulated in Table IV. The significant parameter values for these fitted spectra are $z_{t} = 6$~kpc, $L = 18$~kpc,  $D_0 =1.8\times 10^{29}$~ $\rm{cm^2/s}$ and $B_0^{\rm{turbulent}} =6.62 \times 10^{-6}$ ~ Gauss.}
\end{figure*}

\begin{table}[!ht]
\caption{\label{tab:tab3} Various models and parameter values selected in \texttt{DRAGON} code to fit CR spectra shown in Figs.~6(a,b,c,d) are listed here. The density profiles of Case~\#3 have been selected for the study. Significant changes (in comparison to Table~\ref{tab:tab1}) are shown in bold text. }
\begin{ruledtabular}
\begin{tabular}{ |p {5 cm}| l | }

Model/Parameter  &   Option/Value \\
 \hline
  $R_{\rm{max}}$																	 			   &     20.0 ~ kpc \\  
  $L$                      																				&   \bf{18.0~ kpc}  \\
  Gas density type																			&   \bf{Case \verb+#3+ } \\
  
  Source Distribution          															 &   Ferriere \\
  Diffusion type																				&        Exp  (see Eq.~(1))         \\
  $D_0$       																					&   $ \bf{1.8 \times 10^{29}}$~ $\bf{\rm{cm^2/s}}$ \\
  $\rho_0$																						&     3.0~GV            \\
  $\delta$																		                &       0.54          	\\
  $z_t$ 																							&		\bf{6.0~kpc}			\\
  $\eta$   																						&		-0.40		    \\
  $v_A$   																					&			0.0		        \\
  Magnetic field type																&	    Pshirkov			      \\
  $B_0^{\rm{disc}}$ 															&      $2.0 \times 10^{-6}$ ~ Gauss      \\
  $B_0^{\rm{halo}}$            												&    $4.0 \times 10^{-6}$ ~ Gauss                \\
  $B_0^{\rm{turbulent}}$   											 &       $\bf{6.62 \times 10^{-6}}$ ~ \bf{Gauss}             \\
  First injection slope ($\alpha_0$) 									&  2.32         \\
  Position of first break (rigidity)	                  					 &    330 ~GV      \\
  Second injection slope ($\alpha_1$)         						  &      2.20     \\
    
\end{tabular}
\end{ruledtabular}
\end{table}

In this case, we have considered the density profiles of Case~\#3. We follow the same procedure to fit the observed CR data as used in Case~1. In Table~\ref{tab:tab3}, we have listed various models and parameter values used in the \texttt{DRAGON} code to fit the CR spectra along with the chosen density profile. We need $z_{t} = 6$~kpc, $L = 18$~kpc, $D_0 =1.8 \times 10^{29}$~ $\rm{cm^2/s}$ and $B_0^{\rm{turbulent}} =6.62 \times 10^{-6}$ ~ Gauss to fit $^{10}$Be/$^{9}$Be, B/C, p, He, and $\bar{p}$ spectra for Case~\#3. The other parameters remain same as those mentioned in Table~\ref{tab:tab1}.

\section{Summary and conclusions}

In the present work,  our  prime motive is to study the effects of variations in the density profiles of molecular, atomic and ionized hydrogen gas in the Milky Way Galaxy on its height of the halo. For this purpose we considered the model of GMF consistent with the Faraday rotation measurements and synchrotron emissions. We have also fitted the observed cosmic ray data of $^{10}$Be/$^{9}$Be, B/C, proton, helium, and antiproton with our calculated fluxes using the \texttt{DRAGON} code. We discuss about the different components of hydrogen gas and their observational signatures. 

We present the different density profiles as three different cases, namely  Case~\#1, Case~\#2, and  Case~\#3. An outline of such density profiles with their corresponding references are given in Table~I.  The density profiles in different cases, studied here, are characteristically different (especially Case~\#1 and Case~\#2) from each other. Each of the density profiles is constructed on the basis of various realistic observations, hydrodynamical simulations including gas flows and cosmological parameters, and theoretical modelings.  Due to complex nature of the density profiles, it is not straightforward to compare the cases right away. But we can point out a major difference between Case~\#1 and other cases (for Case~\#2 and  Case~\#3 ) and that is the components of hydrogen gas in Case~\#1 have very low contribution in the Galactocentric distance of $0-3$~kpc compared to other cases. Hence, the gas target density for the interaction with primary CRs is low enough which in turn supports the fact that the value of  $z_{t}/D_{0}$, for a fixed $R_{\rm{max}}$, should be increased to fit B/C. This could be the physical explanation for a high $z_{t}/D_{0}$ value ( see Table~II) in Case~1  compared to other lower $z_{t}/D_{0}$  values (see Tables~III and IV). On the other hand, the vertical distributions, for $r > 3$~kpc, for both the density profiles of  Case~\#2 and  Case~\#3 are different.  So, Case~\#3 provides more target gas density than Case~\#2. Hence,  $z_{t}/D_{0}$ value in Case~ 2, for a fixed $R_{\rm{max}}$,  is higher than the value of Case 3.    

In a previous work, the halo height was estimated by using a semi-analytic code with various simplified assumptions \cite{putze10}. In that work, the ISM gas density was considered as $\sim 1~{\rm{cm^{-3}}}$. Later, the estimate of halo height was improved with the help of numerical CR propagation model. The work was done by using Bayesian analysis along with the \texttt{GALPROP} code \footnote{\url{http://galprop.stanford.edu}} which includes  density distributions of molecular, atomic and ionized hydrogen gas \cite{trotta11}. The estimation of halo height has also been revisited by using \texttt{DRAGON} code \cite{bernardo13}. The gas density profile in \texttt{DRAGON} code has almost similar form as \texttt{GALPROP}. In our analysis, we implement different types of physically possible density profiles of hydrogen gas in the \texttt{DRAGON} code. The density profiles of the  innermost region of the Galaxy ($\lesssim 3$~kpc) have also been included in Case\#2 and Case\#3.      

 The halo height can be treated as a corner stone of modern astroparticle physics as the knowledge of it is very useful for both the conventional CR astrophysics and the indirect search of dark matter. We find the required value of halo height ($z_{t}$) is in the range of 2-6 kpc to fit the CR data for the density profiles considered in our work. Our detailed quantitative analysis shows the density profiles of our Galaxy available from different observations and hydrodynamical simulations yield significantly different estimates of the halo height. Hence it is important to have more observational and theoretical studies in future to determine the density profiles uniquely.

\section*{Acknowledgment}
We would like to thank the anonymous referee for useful comments and suggestions. We also thank K. Ferriere for useful discussions.

\bibliography{main}

\end{document}